\documentstyle[11pt,aaspp4,flushrt]{article}

\lefthead{Aloy et al.}
\righthead{GENESIS: A high--resolution code for
3D relativistic hydrodynamics}
\def\be{\begin{equation}}
\def\ee{\end{equation}}
\def\etal{et al.}
\newcommand\cref[1]{(\ref{#1})}
\begin{document}
\singlespace
\title{GENESIS: \\ A high--resolution code for
3D relativistic hydrodynamics}

\author{M.A. Aloy, J.M$^{\underline{\mbox{a}}}$ Ib\'a\~nez, 
J.M$^{\underline{\mbox{a}}}$ Mart\'{\i}}
\affil{Departamento de Astronom\'{\i}a y Astrof\'{\i}sica\\
Universidad de Valencia, 46100 Burjassot (Valencia), Spain}

\and

\author{ E. M\"uller}
\affil{Max-Planck-Institut f\"ur Astrophysik\\
Karl-Schwarzschild-Str. 1, 85748 Garching, Germany}

\begin{abstract}
The main features of a three dimensional, high--resolution special
relativistic hydro code based on relativistic Riemann solvers are
described. The capabilities and performance of the code are discussed.
In particular, we present the results of extensive test calculations
which demonstrate that the code can accurately and efficiently handle
strong shocks in three spatial dimensions. Results of the performance
of the code on single and multi-processor machines are given.
Simulations (in double precision) with $\le 7 \,\, 10^6$ computational
cells require less than 1\,Gb of RAM memory and $\approx 7 \,\,
10^{-5}$ CPU seconds per zone and time step (on a SCI Cray--Origin
2000 with a R10000 processor). Currently, a version of the numerical
code is under development, which is suited for massively parallel
computers with distributed memory architecture (like, e.g., Cray T3E).
\end{abstract}


\section{Introduction } \label{intro}

  Numerical relativistic hydrodynamics (RHD) has experienced an
important step forward in recent years when modern high--resolution
shock--capturing (HRSC) techniques began to be applied to solve the
equations of RHD in conservation form. Prior to the advent of HRSC
techniques the field was dominated for more than one decade by Wilson
(1979)'s approach to relativistic hydrodynamics.
This approach relies on the use of artificial viscosity in order to
handle the discontinuities (shocks, contact discontinuities, etc.)
that may appear in the flow numerically. However, techniques based on
artificial viscosity are prone to severe numerical difficulties when
simulating ultrarelativistic flows (see, e.g., Centrella \& Wilson
1984). Using modern HRSC techniques instead, these difficulties are
overcome (see, e.g., Donat et al. 1998) allowing one to simulate
challenging relativistic astrophysical phenomena like, e.g.,
relativistic jets or gamma-ray bursts (GRB hereafter).

  In astrophysical jets flow velocities as large as $99.5\%$ of the
speed of light (Lorentz factors $>10$) are required -- according to
the nowadays accepted standard model -- to explain the apparent
superluminal motion observed at parsec scales in many jets of
extragalactic radio sources associated to active galactic nuclei.
Similar arguments applied to the galactic superluminal sources
GRS1915+105 (Mirabel \& Rodriguez 1994) and GROJ1655--40 (Tingay et
al. 1995) allow one to infer intrinsic velocities of $\approx 0.9c$ in
the jets of these sources. Further independent indication of highly
relativistic speeds can be inferred from the intraday variability
occurring in more than a quarter of all compact extragalactic radio
sources (Krichbaum, Quirrenbach \& Witzel 1992). If the observed
intraday radio variability is intrinsec and results from incoherent
synchrotron radiation (according to Begelman, Rees \& Sikora 1994),
the associated jets must have bulk Lorentz factors in the range $\sim$
30 -- 100.

  With exception of the remarkable work of van Putten (1993, 1996),
who used pseudo spectral techniques to solve the equations of
relativistic magnetohydrodynamics, numerical simulations of
relativistic jets started soon after the first multidimensional
relativistic HRSC codes had been developed (Mart\'{\i}, M\"uller \&
Ib\'a\~nez 1994, Duncan \& Hughes 1994, Mart\'{\i} et al. 1995). Since
then many different aspects of relativistic jets have been
investigated (see, e.g., Mart\'{\i} 1997 for a recent review). The
morphology, dynamics and propagation properties of relativistic jets
have been analysed in Mart\'{\i} et al. (1997).  Komissarov \& Falle
(1997) investigated the long--term evolution of relativistic
jets. First simulations of superluminal sources combining relativistic
hydrodynamics and synchrotron radiation transfer at parsec scales have
been performed by G\'omez et al. (1995, 1997) and by Mioduszewski,
Hughes \& Duncan (1997) and Komissarov \& Falle (1997).  From their
simulations these authors inferred that the observations of such
sources can be explained in terms of travelling perturbations in
steady relativistic jets.

In the last two years further progress was achieved by simulating
relativistic jets in three spatial dimensions and by incorporating
magnetic fields (Koide, Nishikawa \& Mutel 1996; Nishikawa et
al. 1997; Koide 1997).  However, instead of using fluxes obtained by
solving Riemann problems at zone interfaces, Koide and collaborators
code rely on the addition of nonlinear dissipation terms to their
Lax--Wendroff scheme to stabilise the code across discontinuities.
This stabilisation method was originally proposed by Davis (1984), who
applied it successfully to the equations of classical hydrodynamics.
The method is robust and simple as no detailed characteristic
information is needed. Koide and collaborators did simulate the
evolution of the jet only for a very brief period of time. This fact
and the coarse grid zoning used in their simulations, however,
prevented them from studying genuine 3D effects in relativistic jets
in any detail.  On the other hand, the relative smallness of the beam
flow Lorentz factor (4.56, beam speed $\approx 0.98$) assumed in their
simulations does not allow for a comparison with Riemann-solver-based
HRSC methods in the ultrarelativistic limit.

An astrophysical phenomenon which also involves flows with velocities
very close to the speed of light are gamma-ray bursts (GRB). Although
known observationally since over 30 years, their nature and their
distance (``local'' or ``cosmological'') is still a matter of
controversial debate (Fishman \& Meegan 1995; M\'ez\'aros 1995; Piran
1997).  In order to explain the energies released in a GRB various
catastrophic collapse events have been proposed including
neutron-star/neutron-star mergers (Pacy\'nski 1986; Goodman 1986;
Eichler \etal 1989), neutron-star/black-hole mergers (Mochkovitch
\etal 1993) collapsars (Woosley 1993) and hypernovae (Pacy\'nski
1998). These models all rely on a common engine, namely a stellar mass
black hole which accretes several solar masses of matter from a disk
(formed during a merger or by a non-spherical collapse) at a rate of
$\sim 1\,\mbox{M}_{\odot}\mbox{s}^{-1}$ (Popham, Woosley \& Fryer 1998). 
A fraction of
the gravitational binding energy released by accretion is converted
into neutrino and anti-neutrino pairs, which in turn annihilate into
electron-positron pairs. This creates a pair fireball, which will also
include baryons present in the environment surrounding the black
hole. Provided the baryon load of the fireball is not too large, the
baryons are accelerated together with the e$^+\,$e$^-$ pairs to
ultrarelativistic speeds with Lorentz factors $> 10^2$ (Cavallo \&
Rees 1978; Piran, Shemi \& Narayan 1993). The bulk kinetic energy of
the fireball then is thought to be converted into gamma-rays via
cyclotron radiation and/or inverse Compton processes (see, e.g.,
M\'esz\'aros 1995).

  In the following we describe the main features of a special
relativistic 3D hydrodynamic code, which is based on explicit HRSC
methods, and which is a considerably extended version of the special
relativistic 2D hydrodynamic code developed by Mart\'{\i}, M\"uller \&
Ib\'a\~nez (1994) and by Mart\'{\i} et al. (1995). The code has been
designed modularly which allows one to use different reconstruction
algorithms and Riemann solvers. 
%
As it is the final goal of our work to simulate relativistic jets and
GRBs in three spatial dimensions, the code has successfully been
subjected to an intensive testing in the ultrarelativistic regime (see
Section \cref{ctest}). In particular, GENESIS has successfully passed
the spherical shock reflection test (simulated in 3D Cartesian
coordinates) involving flow Lorentz factors larger than 700 (see \S
\cref{ESSR}).

  The paper is organised as follows. In Section \cref{eqs}, we
introduce the 3D equations of RHD in Cartesian coordinates in
differential and discretized forms. The latter have been implemented
into our 3D RHD code GENESIS. Detailed information about the structure
and the main features of the code is given in Section \cref{struc}.
Several 1D, 2D and 3D relativistic test problems computed with GENESIS
are described in Section \cref{ctest}. The performance of GENESIS on
scalar and multi-processor computers is analysed in Section
\cref{cperf}, and a realistic simulation of a 3D relativistic
astrophysical jet is presented in Section \cref{jet}. A summary of the
paper containing our main conclusions and a discussion of present and
future applications of the code in different astrophysical areas can
be found in Section \cref{conc}. In Appendix~A, we give the spectral
decomposition of the three dimensional system of RHD equations with
explicit expressions for the eigenvalues and the right- and
left-eigenvectors. 
%
Appendix~B contains the explicit formulae for the numerical viscosity
for Marquina's (Donat \& Marquina 1996) Riemann solver, and Appendix~C
describes the explicit algorithm to recover the primitive variables
form the conserved ones.

\section{Equations of RHD in conservation form} \label{eqs}

The evolution of a relativistic perfect fluid is described by five
conserved quantities: rest mass density, $D$, momentum density, ${\bf
S}$, and energy density, $\tau$ (all of them measured in the
laboratory frame and in natural units, i.e., the speed of light $c=1$),
\begin{eqnarray}
 D &=& \rho W \label{D} \\
 S^{j} &=&  \rho h W^2 v^j \quad (j = 1, 2, 3) \label{S} \\
 \tau &=&  \rho h W^{2}-p-\rho W, \label{tau}  
\end{eqnarray}
where the Lorentz factor $W = (1-{\rm v}^{2})^{-1/2}$ and ${\rm v}^{2}
= \delta_{ij} v^i v^j$ 
%
(the Einstein summation convention is used here, and $\delta_{ij}$ is
the Kronecker symbol).
Furthermore, $\rho$ is the rest--mass density, $p$ the pressure and
$h$ the specific enthalpy given by $h = 1 + \varepsilon + p/\rho$ with
$\varepsilon$ being the specific internal energy. The components of
the vector of variables ${\bf w} \equiv (\rho, v^{i}, \varepsilon)^T$
are called {\it primitive} or physical variables.

The relativistic Euler equations form a system of conservation laws
(see, e.g., Font et al. 1994) which can be written in Cartesian
coordinates as
\begin{eqnarray}
 \frac{\partial D}{\partial t} & + & \displaystyle{\sum_{j=1}^{3}}
 \frac{\partial}{\partial x^{j}}(Dv^{j}) = 0 \\
 \frac{\partial S^{i}}{\partial t} & + & \displaystyle{\sum_{j=1}^{3}}
 \frac{\partial}{\partial x^{j}}(S^{i}v^{j}+{\delta}^{ij}p)= 0 
 \:\:\:\:\:\:\:\: \mbox{(i=1, 2, 3)} \\
 \frac{\partial \tau}{\partial t} & + & \displaystyle{\sum_{j=1}^{3}}
 \frac{\partial}{\partial x^{j}}(S^{j}-Dv^{j})= 0     
\end{eqnarray}
or, equivalently, as
\begin{equation}
 \frac {\partial {\bf U}}{\partial t} + \displaystyle{\sum_{j=1}^{3}}
 \frac {\partial {\bf F}^{j}}{\partial x^{j}} = 0 \, , 
 \label{F}
\end{equation}
where the vector of unknowns ${\bf U}$ (i.e., the conserved variables)
is given by
\begin{equation}
 {\bf U} = \left(D, \: S^1, \: S^2, \: S^3, \: {\tau} \right)^{T} \, ,
\label{U}
\end{equation}
and the fluxes are defined by
\begin{eqnarray}
 {\bf F}^{i} & =  & \left(D v^{i}, S^1 v^i + p \delta^{1i},
                          S^2 v^i + p \delta^{2i}, S^3 v^i + p \delta^{3i},
                          S^i - Dv^i \right)^T \, .
\end{eqnarray}

The system \cref{F} of partial differential equations is closed with
an equation of state $p=p(\rho,\varepsilon)$. Anile (1989) has shown
that system \cref{F} is hyperbolic for causal equations of state,
i.e., for those where the local sound speed, $c_s$, defined by
\begin{equation}
 h c_{s}^{2} = \frac{\partial p }{ \partial \rho} + (p/\rho^{2})
 \frac{\partial p }{ \partial \epsilon}, 
 \label{cs}
\end{equation}
satisfies $c_s <1$.

The structure of the characteristic fields corresponding to the 
nonlinear system of conservation laws \cref{F} has explicitly been 
derived in Donat et al. (1998) and is summarised in Appendix~A.

In order to evolve system \cref{F} numerically, one has to discretize
the state vector ${\bf U}$ within computational cells. The temporal
evolution of the state vector is determined by the flux balance across
the zone interfaces of each cell and the contribution of source
terms. Using a method of lines (see, e.g., LeVeque 1991), our
discretization reads
\begin{eqnarray}  \label{RKec}
 \frac{d{\bf U}_{i,j,k}}{dt} & = &
  -\frac{1}{{\Delta}x}\left({\bf \tilde{F}}^x_{i+\frac{1}{2},j,k} - 
    {\bf \tilde{F}}^x_{i-\frac{1}{2},j,k}\right) 
  -\frac{1}{{\Delta}y}\left({\bf \tilde{F}}^y_{i,j+\frac{1}{2},k}-
    {\bf \tilde{F}}^y_{i,j-\frac{1}{2},k}\right) \nonumber \\
  \mbox{} &   &
  -\frac{1}{{\Delta}z}\left({\bf \tilde{F}}^z_{i,j,k+\frac{1}{2}}-
    {\bf \tilde{F}}^z_{i,j,k-\frac{1}{2}}\right) 
   +{\bf S}_{i,j,k} \equiv L ({\bf U})\, ,
\end{eqnarray}
where latin subscripts $i$, $j$ and $k$ refer to the $x$, $y$ and $z$
coordinate direction, respectively. ${\bf U}_{ijk}$ and ${\bf
S}_{i,j,k}$ are the mean values of the state and source vector (if non
zero) in the corresponding three-dimensional cell, while ${\bf
\tilde{F}}^x_{i \pm \frac{1}{2},j,k}$, ${\bf \tilde{F}}^y_{i,j \pm
\frac{1}{2},k}$ and ${\bf \tilde{F}}^z_{i,j,k \pm \frac{1}{2}}$ are
the numerical fluxes at the respective cell interface.
%
Finally, $L({\bf U})$ is a short hand notation of the spatial operator
in our method.

  At this stage, our system of conservation laws is a system of
ordinary differential equations which can be integrated with a large
number of algorithms. We have chosen a multistep Runge--Kutta (RK)
method developed by Shu \& Osher (1988) which can provide second (RK2)
and third (RK3) order in time. The explicit form of the algorithms is
(subindexes $(i,j,k)$ are ommited to clarify the notation):
\begin{enumerate}
    \item Prediction step (common for both RK2 and RK3):
    \begin{equation}
    {\bf U}^{(1)} = {\bf U}^n+{\Delta}t L({\bf U}^n) \label{RKa}
    \end{equation}
    \item Depending on the order do:
        \begin{itemize}
            \item RK2:
               \begin{equation}
               {\bf U}^{n+1} = \frac{1}{\alpha} \left (
                               {\beta}{\bf U}^n+{\bf U}^{(1)}+
                               {\Delta}t L({\bf U}^{(1)}) \right ),
               \end{equation}
            being $\alpha = 2$ and $\beta = 1$.
            \item RK3:
               \begin{eqnarray}
               {\bf U}^{(2)} & = & \frac{1}{\alpha} \left (
                               {\beta}{\bf U}^n+{\bf U}^{(1)}+
                               {\Delta}t L({\bf U}^{(1)}) \right ) \nonumber \\
               {\bf U}^{n+1} & = & \frac{1}{\beta} \left (
                               {\beta}{\bf U}^n+2{\bf U}^{(2)}+
                               2{\Delta}t L({\bf U}^{(2)}) \right ), 
                               \label{RKb}\\
               \end{eqnarray}
            in this case, $\alpha = 4$ and $\beta = 3$.** 
        \end{itemize}
\end{enumerate}

\section{The relativistic hydrodynamic code GENESIS} \label{struc}
\subsection{Code structure}

The special relativistic multidimensional hydrodynamic code GENESIS
described in detail in the following is a 3D extension of the 2D HRSC
hydrodynamic code developed by some of the authors. The 2D code has
been successfully used for the simulation of relativistic jets
(Mart\'{\i} et al. 1994, 1995, 1997; G\'omez et al. 1995, 1997). The
main structural features of Mart\'{\i} et al.'s code has been kept,
but there are important changes in the computational part. Besides the
addition of the third spatial dimension, a large effort has been made
to {\it minimise memory requirements} and to {\it optimise the
performance} of the code as well as to enhance its {\it portability}.

Like its predecessor, GENESIS evolves the equations of RHD in
conservation form using a finite volume approach in Cartesian
coordinates. In accordance with the method of lines, we split the
discretization process in two parts. First, we only discretize the
differential equations in space, i.e. the problem remains continuous
in time. This leads to a system of ordinary differential equations
(ODEs) in time \cref{RKec}. The numerical fluxes between adjacent
cells required for the time integration are obtained by solving the
appropriate 1D Riemann problems along the coordinate directions
(spatial sweeps).  High-order spatial accuracy is achieved by applying
a high-order interpolation procedure in space, while high-order
accuracy in time is obtained by using high-order ODE solvers.

GENESIS integrates the 3D RHD equations on uniform grids in each
spatial direction. In order to have a flexible code GENESIS is
programmed to allow for different boundary conditions, spatial
reconstruction algorithms, Riemann solvers, ODE solvers for the time
integration and external forces. The user selects these options at the
preprocessor level, which reduces the number of {\it if}--clauses
inside the nested 3D loops to a minimum, and thereby maximising the
code's efficiency.

Making the selection at the preprocessing stage has allowed us to
obtain a code, which is independent of a specific (shared memory)
machine architecture.  Hence, it runs on different types of machines
and processors. Up to now, we have tested GENESIS on SGI platforms
(INDY workstations, Power Challenge and Cray-Origin 2000 arrays), on
HP machines (712 workstations and J280 computers), and on a CRAY-JEDI
multiprocessor system. As a next step we plan to port GENESIS on a
CRAY-T3E massively parallel computer.

\begin{figure}
\plotone{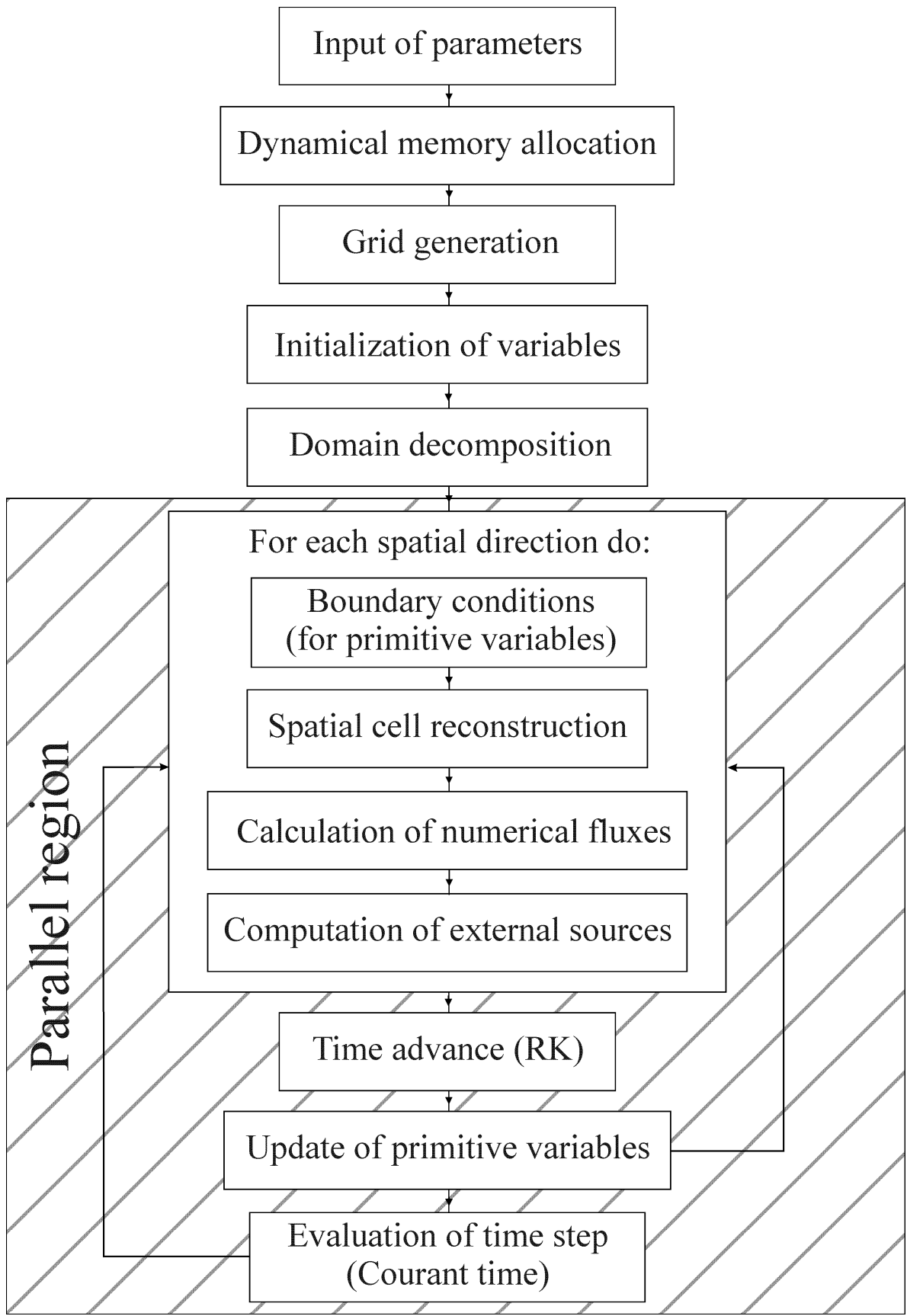}
\caption{Flow diagram of GENESIS. \label{flowdiag}}
\end{figure}

\placefigure{flowdiag}

The flow diagram of GENESIS is shown in figure \ref{flowdiag}. Details
of the major components of GENESIS are discussed in the following
subsections.

\subsection{Memory requirements}

  The current version of GENESIS, which is written in FORTRAN~90 has
the capability of allocating memory {\it dynamically}, i.e. the number
of computational cells can be chosen at run time. Reducing the RAM
requirements of a 3D hydrodynamic code is obviously crucial. In
GENESIS multidimensional variables are responsible for about $99\%$ of
the code's memory requirement. Thus, the number of these 3D arrays has
to be kept at the absolute minimum possible. In its present version,
GENESIS only requires three sets of five 3D arrays each, consisting of
one set of conserved variables at the begining of each time level
(${\bf U}^n$), another set of primitive variables and a third set of
scratch variables (${\bf \tilde U}$).
%
The time integration scheme (eqs.  \ref{RKa}--\ref{RKb}) which results
in the updated values of the conserved variables at the next time
level (${\bf U}^{n+1}$) then reads:

\begin{enumerate}
    \item Prediction step (common for RK2 and RK3):
    \begin{eqnarray*}
    {\bf \tilde U} = {\bf U}^n+{\Delta}t L({\bf U}^n)
    \end{eqnarray*}
%
    \item Depending on the order of accuracy of the time integration scheme do:
        \begin{itemize}
            \item [RK2:]
               \begin{eqnarray*}
               {\bf \tilde U} & = & {\bf \tilde U}+{\Delta}t L({\bf \tilde U}), \\
               {\bf U}^{n+1} & = & \frac{1}{\alpha} \left (
                               {\beta}{\bf U}^n+{\bf \tilde U} \right ),
               \end{eqnarray*}
%
            with $\alpha = 2$ and $\beta = 1$, or
            \item [RK3:]
               \begin{eqnarray*}
               {\bf \tilde U} & = & {\bf \tilde U}+{\Delta}t L({\bf \tilde U}), \\
               {\bf \tilde U} & = & \frac{1}{\alpha} \left (
                               {\beta}{\bf U}^n+{\bf \tilde U} \right), \\
               {\bf \tilde U} & = & {\bf \tilde U}+{\Delta}t L({\bf \tilde U}), \\                 
               {\bf U}^{n+1} & = & \frac{1}{\beta} \left (
                               {\beta}{\bf U}^n+2{\bf \tilde U} \right ),
               \end{eqnarray*}
%
            with $\alpha = 4$ and $\beta = 3$. 
        \end{itemize}
\end{enumerate}

Quantities like entropy, internal energy, sound speed or Lorentz
factor are implemented as FORTRAN scalars. Consequently, GENESIS needs
about 1 Gbyte of RAM memory to handle a grid of $100 \times 100 \times
720$ (in double precision).

\subsection{Domain decomposition}

The technique of domain decomposition is used to optimise the
parallelization of the code and to guarantee its performance in real
applications, too. It is also the first step towards the development
of a parallel version of GENESIS which runs efficiently on parallel
computers with distributed memory.

\begin{figure}
\plotone{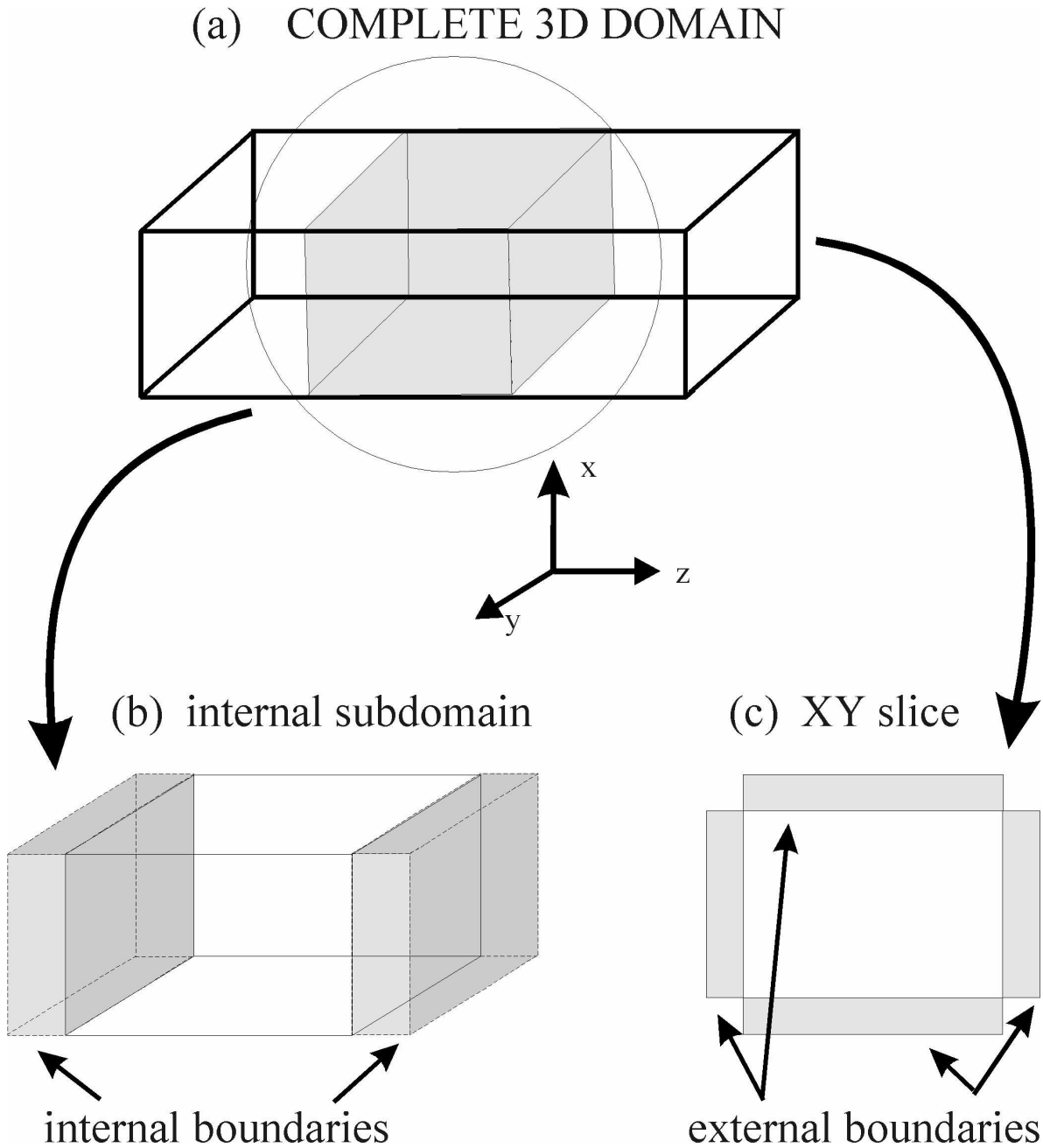}
\caption{(a) Complete three dimensional computational domain, showing a typical
subdomain (in grey). (b) Zoom of the previous subdomain including its internal 
boundaries. These regions overlap with contiguous subdomains. (c) Cut through 
the computational grid along the $X$-$Y$ plane displaying the external 
boundaries.}
\label{domdec}
\end{figure}

The physical domain is split along {\it one} arbitrary spatial
direction (z, in the present version) in a set of subdomains
(i.e. {\it slices}, see Fig.~\ref{domdec}a) of similar computational
load.  The subdomains are then distributed across
processors. Numerical fluxes at subdomain boundaries are calculated by
providing the appropriate internal and external boundary conditions (see
Fig.~\ref{domdec}b,c, respectively, and \S~\ref{bound}).

\placefigure{domdec}

\subsection{Boundary conditions}
\label{bound}
  The computational grid is extended in each coordinate in positive and
negative direction by four so-called {\it ghost} zones, which provide
a convenient way to implement different types of boundary
conditions. These boundary conditions have to be provided in each
spatial sweep for all primitive variables.  In GENESIS several types
of boundary conditions are available including reflecting, inflow,
outflow, time-dependent and analytically prescribed boundary
conditions.

  Flow conditions at subdomain boundaries must be provided, too, in
order to calculate numerical fluxes at subdomain interfaces. Hence,
subdomains are also enlarged by four ghost zones in each coordinate
direction. Note that these ghost zones do overlap with adjacent
subdomains (see Fig. \ref{domdec}).  The internal boundary conditions
in these overlapping regions are defined by copying the corresponding
values of the respective adjacent subdomain.
For $NS$ subdomains and $NX \times NY \times NZ$ computational zones
the number of overlapping cells is $(4+4) \times (NS-1) \times NX
\times NY$, i.e. the fraction of overlapping cells is $8 \times (NS-1)
/ NZ$. Hence, for $NS = 16$ and $NZ = 1000$ (typical of a jet
simulation) the fraction of overlapping cells is about 12\%.

\subsection{Spatial reconstruction}

  In order to improve the spatial accuracy of the code, we interpolate
the values of the pressure, the proper rest--mass density and the
spatial components of the four--velocity ($Wv^i$) within computational
cells. These reconstructed variables are afterwards used to compute
the numerical fluxes. Because of the monotonicity of the
reconstruction procedures (see below) used in GENESIS, the occurrence
of unphysical (i.e., negative) values in the reconstructed profiles of
pressure and density are avoided. In addition, reconstructing the
spatial components of the four-velocity with monotonic schemes, also
prevents the occurrence of unphysical values of the flow velocity,
i.e., the flow velocity always remains smaller than the speed of light
even in multidimensional calculations.

  GENESIS provides, at the preprocessing level, four different types of
reconstruction schemes: piecewise constant, linear using the minmod
function of Van Leer (1979), parabolic using the piecewise parabolic
method, PPM, of Colella \& Woodward (1984; see also Mart\'{\i} \&
M\"uller 1996) or hyperbolic using the piecewise hyperbolic method,
PHM, of Marquina (1994).

\subsection{Source terms}

  Gravity, local radiative processes, etc., are coupled with
hydrodynamics through terms on the right hand side of the RHD
equations (i.e., via the source terms, ${\bf S}_{i,j,k}$, in
Eq.~\cref{RKec}). GENESIS integrates such terms assuming piecewise
constant profiles for the source functions.

\subsection{Computation of the numerical fluxes} \label{ARS}

  In this paper we use a variant of Marquina's flux formula (see Donat
\& Marquina 1996) which has already been shown to work properly in the
simulation of relativistic jets in 2D (Mart\'{\i} et al. 1997).

  The approach followed by Donat \& Marquina (1996) relies on the
extension of the entropy--satisfying scalar numerical flux of Shu \&
Osher (1989) to hyperbolic systems of conservation laws. Given the
spectral decomposition of the RHD equations (see Appendix~A), the
implementation of Marquina's scheme is straightforward.  

  The original Marquina's algorithm computes the contribution to the
numerical viscosity of each characteristic field in a different way
depending on whether the corresponding eigenvalue (characteristic
speed) does change its sign between the left and right states or
whether it does not. However, instead of using the original algorithm,
we only consider that part which corresponds to characteristic speeds
changing their signs between the left and right states of every
numerical interface. The modified algorithm has a larger numerical
viscosity, but it is more stable and does not involve any {\it
if}-clause. Hence, it can easily be vectorized.

  In the 2D version used in Mart\'{\i} et al. (1997) the left
eigenvectors of the Jacobians are calculated numerically by inverting
the matrix of right eigenvectors. In GENESIS we use the analytical
expressions for the left eigenvectors, which allow one to simplify the
computation of the numerical viscosity terms.

  The explicit expressions for the numerical fluxes (${\bf\tilde{F}}^i$, 
$i \in {x,y,z}$ in Eq.~\ref{RKec}) as a function of the local (reconstructed) 
primitive and conserved variables are given in Appendix~B.
Besides its influence on the efficiency of the code, the use of
explicit expressions for the left eigenvectors also leads to
analytical cancellations in the computation of the numerical viscosity
causing a damping of the growth of round-off errors and an improvement
of the overall accuracy of the code. Previous versions of GENESIS, in
which numerical fluxes were calculated without the use of analytical
expressions, suffered from a growth of round-off errors due to the
large number of operations involved and due to the finite precision of
floating point arithmetics. This growth of errors manifests itself in
a gradual loss of symmetry in initially perfectly symmetric
problems. Our experience shows that the analytical manipulation of the
expressions of the numerical flux together with their appropriate
symmetrization (i.e., using commutating formulas for the components of
the velocity parallel to cell interfaces) allows one to achieve a
perfect numerical symmetry (see \S\ref{symmetry} and \S\ref{RSSR}).

\subsection{Time advance and time step computation}

  Time integration is carried out by two different total variation
diminishing RK methods developed in Shu \& Osher (1988). The user can
choose, at preprocessing level, between the RK2 and RK3 algorithm (see
eqs.  \ref{RKa}--\ref{RKb}). Results of similar quality can be
obtained either with the RK3 algorithm or with RK2 using smaller time
steps. Nevertheless, for a given time step, the computational cost of
RK3 is about a factor 1.5 larger than that of RK2.

  As in any explicit hydrodynamic code, time steps are limited for
stability reasons by the Courant-Friedrichs-Levy (CFL) condition,
which is computed using the characteristic speeds. At the end of each
time step the size of the new time step is determined as the minimum
of the time steps of all subdomains. This requires a global operation
across all subdomains. Experience has shown that acceptable CFL
numbers lie in the interval $[0.1, 0.8]$. CFL numbers larger than
$0.8$ can lead to post shock oscillations.

\subsection{Recovering primitive variables}

  The solution of the Riemann problem requires knowledge of the value of
the pressure and its thermodynamic derivatives. Given the functional
dependence between conserved and primitive variables (see
eq.~\cref{S}), the recovering procedure can not be formulated in
closed form. Instead a kind of iterative method must be used, which is
very time consuming. Hence, usage of the recovering procedure should
be reduced to the absolute minimum. Therefore, primitive variables are
consistently updated from the mean values of the conserved variables
after each Runge-Kutta step and their values are stored in a set of 3D
arrays.

  Our approach is the same as that of Mart\'{\i}, Ib\'a\~nez \& Miralles
(1991) and that of Mart\'{\i} et al. (1997). Its explicit form can be
found in Appendix~C (see also Mart\'{\i} \& M\"uller 1996). The iterative 
recovering procedure is based on a second order accurate Newton-Raphson method
to solve an implicit equation for the pressure.

  In zones where the flow conditions change smoothly the typical number
of iterations ranges from 1 to 3 when a relative accuracy of
$10^{-10}$ is requested. There exist zones, however, inside shocks or
near strong gradients, where the number of iterations required is
larger depending on the strength of the shock or the steepness of the
gradient. For example, in the shock reflection test in 3D, the shock
zone needs about 4 to 8 iterations.

\subsection{Some notes on code structure} \label{symmetry}

  We have taken special care in designing a numerical code that
accurately preserves any symmetries present in the initial data. This
is an important point for a code aimed to study, for example, the stability and 
long term evolution of initially axisymmetric jets.

There exist two potential sources of numerical {\it asymmetries} in
our code, both of them are related to the fact that floating point
arithmetics is not associative. One cause of asymmetries is due to the
computation of numerical fluxes in spatial sweeps, which violates what
we call henceforth {\it sweep--level symmetry} (SLS).  In order to
guarantee SLS the expressions by which the numerical fluxes are
evaluated have been symmetrized (see \S\ref{ARS}).

A second source of (numerically caused) asymmetry arises specifically
in 3D codes using directional splitting. It can only be avoided, if
the code has a property which we call {\it sweep--coherence symmetry}
(SCS).  It refers to the symmetry of the integration algorithm with
respect to the order in which the 1D-sweeps are performed. This
symmetry property of the algorithm becomes crucial if an initially
spherically symmetric state is considered.  We found that its initial
symmetry is lost unless special care is taken in the calculation of
the Lorentz factor (in the numerical flux routine), which involves the
summation of the squares of the three velocity components.  To
guarantee a perfect sweep--coherence symmetry of the algorithm the
addition of the vector components has to be performed in a cyclic
manner, i.e. in the X--sweep the components are summed up in ${x,y,z}$
order, in the Y--sweep in ${y,z,x}$ order, and finally in the Z--sweep
in ${z,x,y}$ order. Due to the stochastic nature of round-off errors,
a violation of the sweep--coherence symmetry manifests itself only in
the last few significant digits of the state variables, if the number
of time steps is not too large (less than about 3000; see section
\S\ref{ESSR}).

Given that round--off errors grow sufficiently slow and that they do
not interact with the truncation errors due to the finite difference
scheme (which can render the scheme {\it unstable}), GENESIS does keep
the symmetry of an initial state at an acceptable level. We have also
tried to develop a version of GENESIS with a perfect 3D symmetry
(limited by the Cartesian discretization). For this purpose, we
applied the {\it extended partial precision} technique in the
computation of expressions in which the associative property should be
satisfied. The procedure was successful, but increased the total
computational costs by more than 30$\%$.  All the results presented in
the following have been obtained without making use of such a
technique.

\section{Code Testing} \label{ctest}

The capabilities of GENESIS to solve problems in special relativistic
hydrodynamics are checked by means of three tests calculations that
involve strong shocks and a wide range of flow Lorentz factors. In
these test runs an ideal gas equation of state with an adiabatic exponent
$\gamma$ has been used. All results presented in this section have
been obtained with the PPM reconstruction procedure and the
relativistic Riemann solver based on Marquina's flux formula (see
previous section for details).

\subsection{Mildly Relativistic Riemann Problem (MRRP)}
    
In the first test we consider the time evolution of an initial
discontinuous state of a fluid at rest. The initial state is given by
$\rho_L=10$, $\epsilon_L=2$, $v_L=0$, $\gamma_L=5/3$, $\rho_R=1$,
$\epsilon_R=10^{-6}$, $v_R=0$ and $\gamma_R=5/3$, where the subscript
$L$ ($R$) denotes the state to the left (right) of the initial
discontinuity. This test problem has been considered by several
authors in the past (in 1D by Hawley, Smarr \& Wilson 1984, Schneider
et al. 1993, Mart\'{\i} \& M\"uller 1996, Wen, Panaitescu \& Laguna
1997; in 2D by Mart\'{\i} et al. 1997). It involves the formation of
an intermediate state bounded by a shock wave propagating to the right
and a transonic rarefaction propagating to the left. The fluid in the
intermediate state moves at a mildly relativistic speed ($v=0.72c$) to
the right. Flow particles accumulate in a dense shell behind the shock
wave compressing the fluid by a factor of 5 and heating it up to
values of the internal energy much larger than the rest-mass energy.
Hence, the fluid is extremely relativistic from a thermodynamical
point of view, but only mildly relativistic dynamically.

To change this intrinsically one dimensional test problem into a
multidimensional one we have rotated the initial discontinuity (normal
to the x-axis) by an angle of $45^o$ around the y-axis, and then again
by an angle of $45^o$ around the z-axis. Gas states $L$ and $R$ are
placed within a cube of major diagonal equal to 1 that constitutes the
3D numerical grid.

\begin{figure}
\plotone{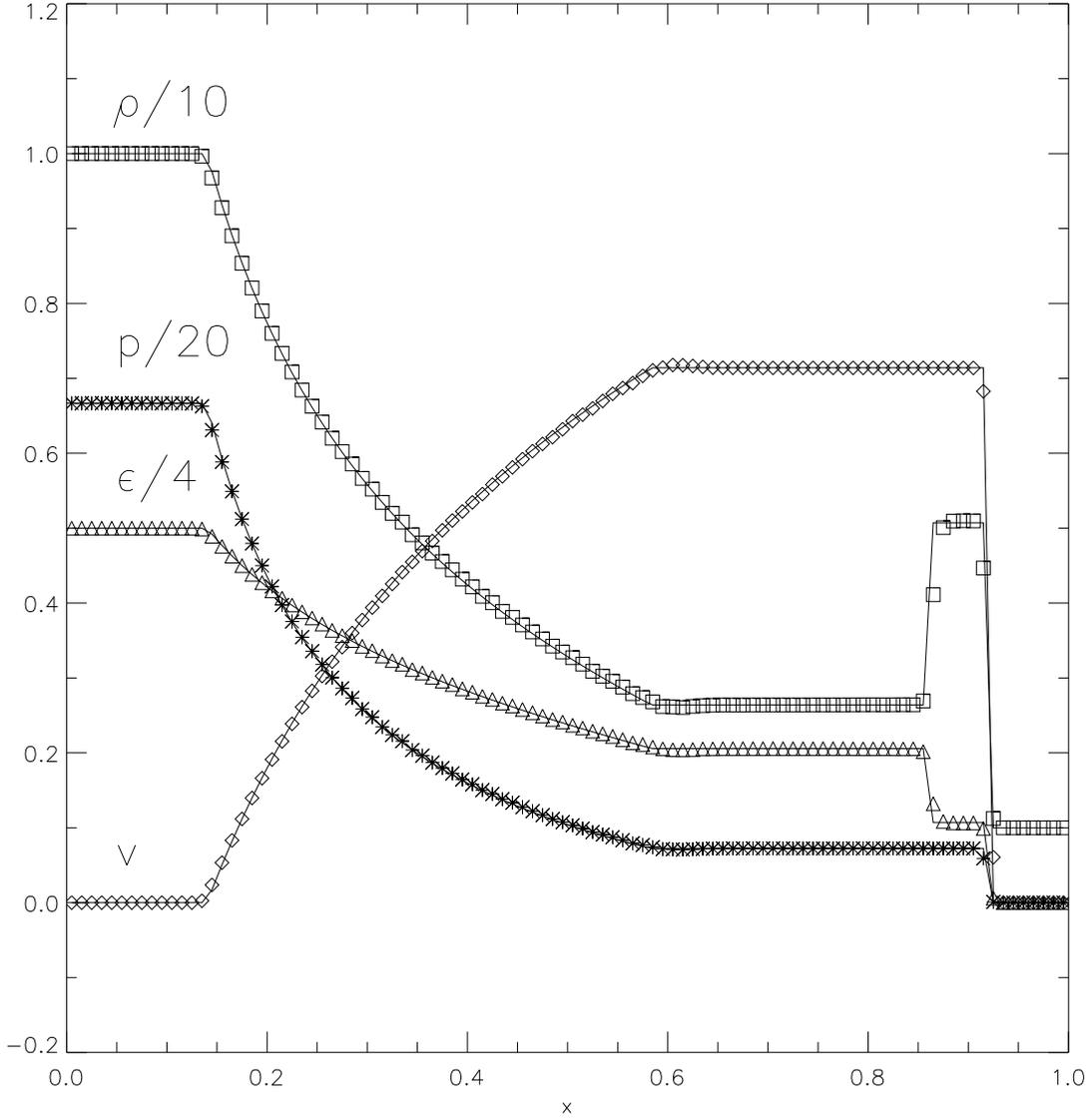}
\caption{Numerical and exact solution of the mildly relativistic
Riemann test problem (MRRP) described in the text after 0.5 time
units. The computed one--dimensional distributions of proper
rest--mass density, pressure, specific internal energy and flow
velocity are shown, in normalized units, with discrete symbols.
Continuous lines depict the corresponding exact solution. The
simulation was performed on a grid of $100^3$ zones. The CFL number
was set equal to 0.6 and a second--order Runge-Kutta was used for time
integration. \label{fig_gsch}}
\end{figure}

The analytical solution to this test problem can be found in
Mart\'{\i} \& M\"uller (1994). Our analysis is restricted to the flow
conditions along the major diagonal of the numerical grid, which is
normal to the initial discontinuity. Figure \ref{fig_gsch} shows the
solution along the major diagonal at time $t=0.5$.  The shock is
captured in two to three zones in accordance with the capabilities of
HRSC methods. The transonic rarefaction has a smooth profile across
the sonic point located at $x=0.5$, and exhibits sharp corners.  The
contact discontinuity is spread out over roughly three zones.

\placetable{tab_err1.schn}

\placefigure{fig_gsch}

The absolute global errors (in $L_1$ norm given by
${\epsilon}_{\mbox{\small abs}} = \sum_{i,j,k}\, \left| {\bf
w}_{i,j,k}^n - {\bf w}({\bf x}_{i,j,k},t_n) \right|\, \Delta x_i
\Delta y_j \Delta z_k$, where ${\bf w}_{i,j,k}^n$ and ${\bf w}({\bf
x}_{i,j,k},t_n)$ are the numerical and exact solution, respectively)
of pressure, density and velocity are given in Table
\ref{tab_err1.schn} for different grid resolutions at $t=0.5$. Table
\ref{tab_err1.schn} implies a convergence rate of slightly less than 1
when comparing the errors obtained on the coarsest ($40^3$) and the
largest ($150^3$) grid. This behaviour is expected for
multidimensional problems involving discontinuities (see, e.g. LeVeque
1991).

\subsection{Relativistic Planar Shock Reflection (RPSR)} \label{RSSR}
  
This 1D test problem involves the propagation of a strong shock wave
generated when two cold gases, moving at relativistic speeds in
opposite directions, collide. The problem has been considered as a
test for almost any new relativistic hydrodynamic code (Centrella \&
Wilson 1984; Hawley, Smarr \& Wilson 1984; Mart\'{\i} \& M\"uller
1994; Eulderink \& Mellema 1994; Falle \& Komissarov 1996).

After the collision of the two gases, two shock waves are created in
the plane of symmetry of the physical domain propagating in opposite
directions. The inflowing gas is heated in the shocks and comes to a
rest.  The exact solution of this Riemann problem was obtained by
Blandford \& McKee (1976).

The initial data are $\rho_L = 1$, $\epsilon_L = 2.29\,10^{-5}$, $v_L
= v_i$, $\rho_R = 1$, $\epsilon_R = 2.29\,10^{-5}$ and $v_R = -v_i$,
where $v_i$ is the inflow velocity of the colliding gas.

\begin{figure}
\plotone{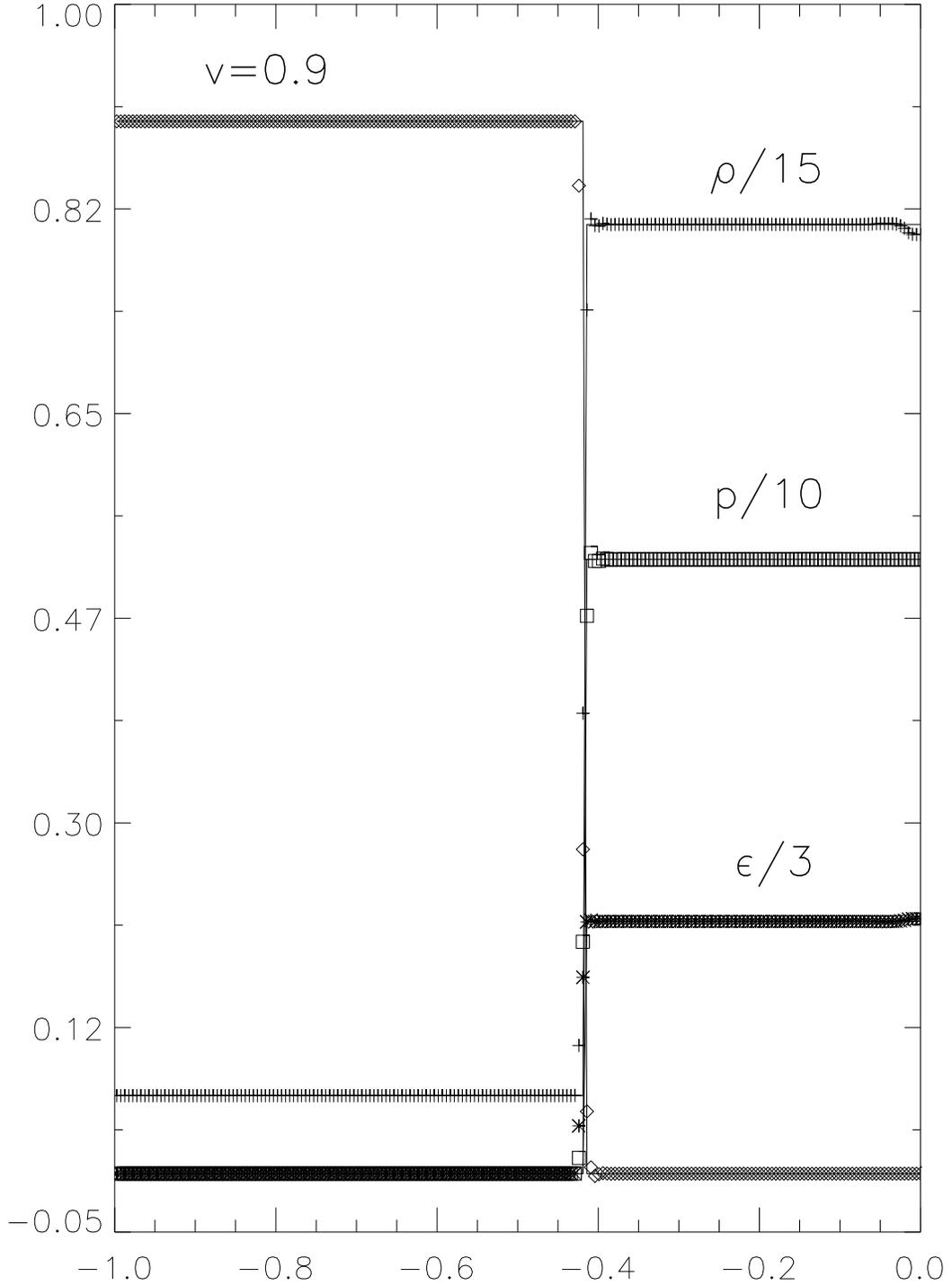}
\caption{Numerical and exact solution of the relativistic planar shock
reflection problem (RPSR) described in the text after 2.0 time
units. The computed distributions of proper rest--mass density,
pressure, specific internal energy and flow velocity are shown, in
normalized units, with discrete symbols, for an inflow velocity of the
colliding gases equal $v_i=0.9$.  Continuous lines depict the
corresponding exact solution. The simulation was performed on a grid
of $401$ zones spanning the interval $[-1,1]$ with both gases
colliding in the middle of the grid at $x=0$. Only the left half of
the grid is shown. The CFL number was set equal to 0.3 and a
second--order Runge--Kutta was used for time integration.
\label{fig_RSSR}}
\end{figure}

\placefigure{fig_RSSR}

Figure \ref{fig_RSSR} shows the numerical solution at $t=2.0$ on the
left half of a grid having a total of 401 zones. The results obtained
in the right half of the grid are strictly symmetric with respect to
the collision point ($x=0$), i.e., the sweep--level symmetry 
(SLS; see section \S\ref{symmetry}) is exactly fulfilled. Near $x=0$,
the numerical solution shows small errors (of the same order as the
mean error in the post-shock state, $0.3 \%$) which are due to the
{\it wall heating} phenomenon (Noh 1987) characterised by an
overshooting of the internal specific energy and an undershooting of
the proper rest-mass density.

In Table \ref{tab_err1.rssr} we give the global absolute errors ($L_1$
norm) of the primitive variables for different grids at $t=2.0$ and
for an inflow velocity $v_i = 0.999c$. We find a convergence rate
about equal to one (see columns 5-7) for all variables.

\placetable{tab_err1.rssr}

We can use this test problem to check the robustness of GENESIS in the
ultrarelativistic regime. To simplify notation, we define the quantity
$\nu = 1 - v_i$, which tends to zero when $v_i$ tends to one. Table
\ref{tab_err4.rssr} contains the relative global errors of the
primitive variables at $t=2.0$ for a set of calculations performed on
a grid of 401 zones, where we have varied $\nu$ from $10^{-1}$ to
$10^{-11}$. The latter value corresponds to a Lorentz factor $W = 2.24
\times 10^{5}$. The relative error of the primitive variables shows a
weak dependence on the inflow velocity. It never exceeds $3.5\%$ and
for $\nu \ge 10^{-9}$ it is smaller than $1\%$.

\placetable{tab_err4.rssr}

\begin{figure}
\plotone{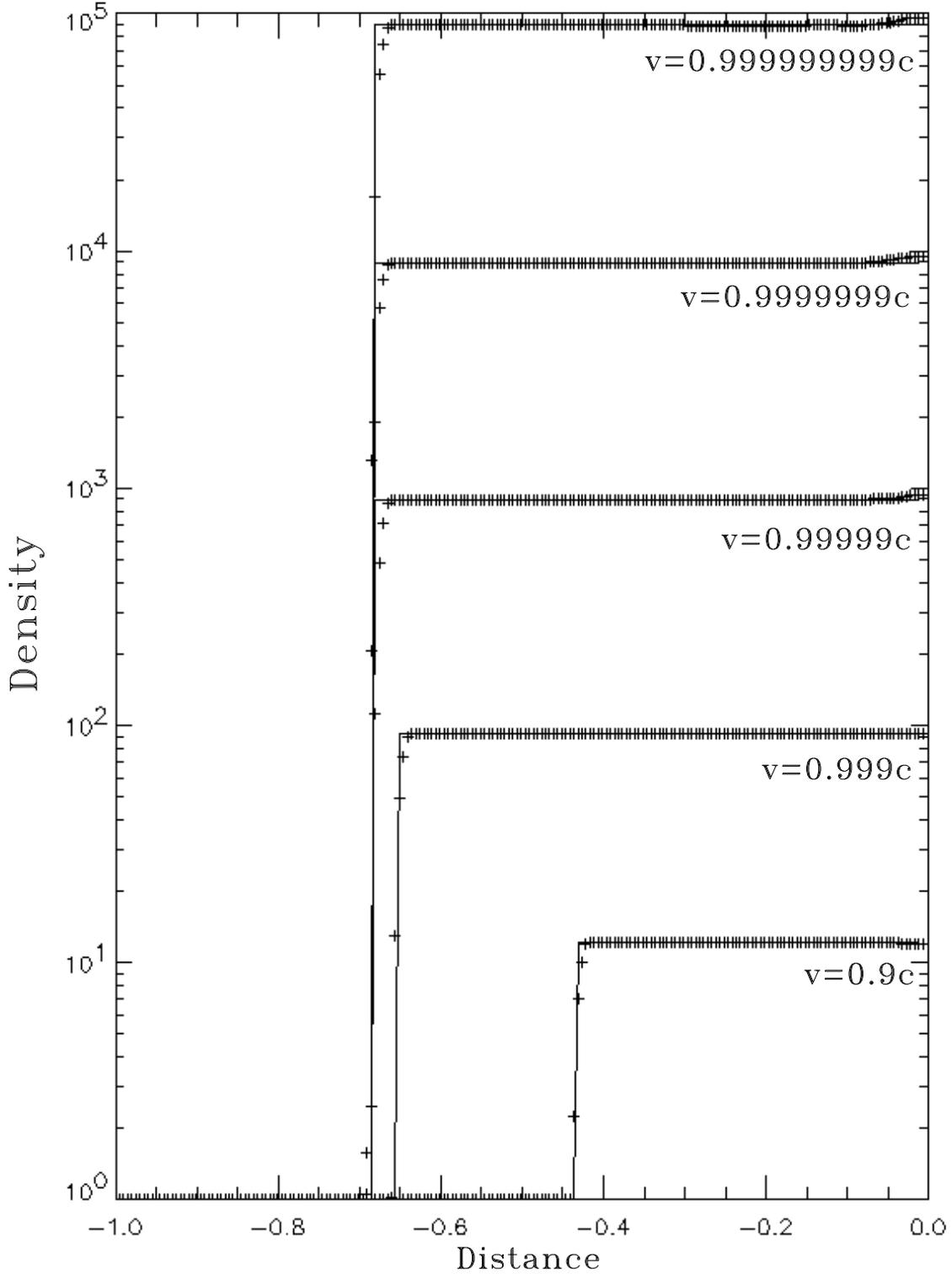}
\caption{Density jump (in logarithmic scale) for different inflow
velocities in the relativistic planar shock reflection problem (RPSR),
over an equally spaced grid of $401$ zones at $t=2.0$. As in the
previous Figure, only the left half of the grid is shown. Solid lines
represent the exact solution while symbols refer to numerical
values. A third--order Runge--Kutta was used for time integration.
\label{fig_RSSR_m}}
\end{figure}

The PPM parameters (see Colella \& Woodward 1984) have been tuned to
minimize the number of zones within the shock without introducing
unacceptable numerical post--shock oscillations. Fig. \ref{fig_RSSR_m}
demonstrates that there are no numerical post-shock oscillations for
$\nu \leq 10^{-5}$ when the shock is captured by 2 to 3 zones.

\placefigure{fig_RSSR_m}

\subsection{Relativistic Spherical Shock Reflection (RSSR)} \label{ESSR}

\begin{figure}
\plotone{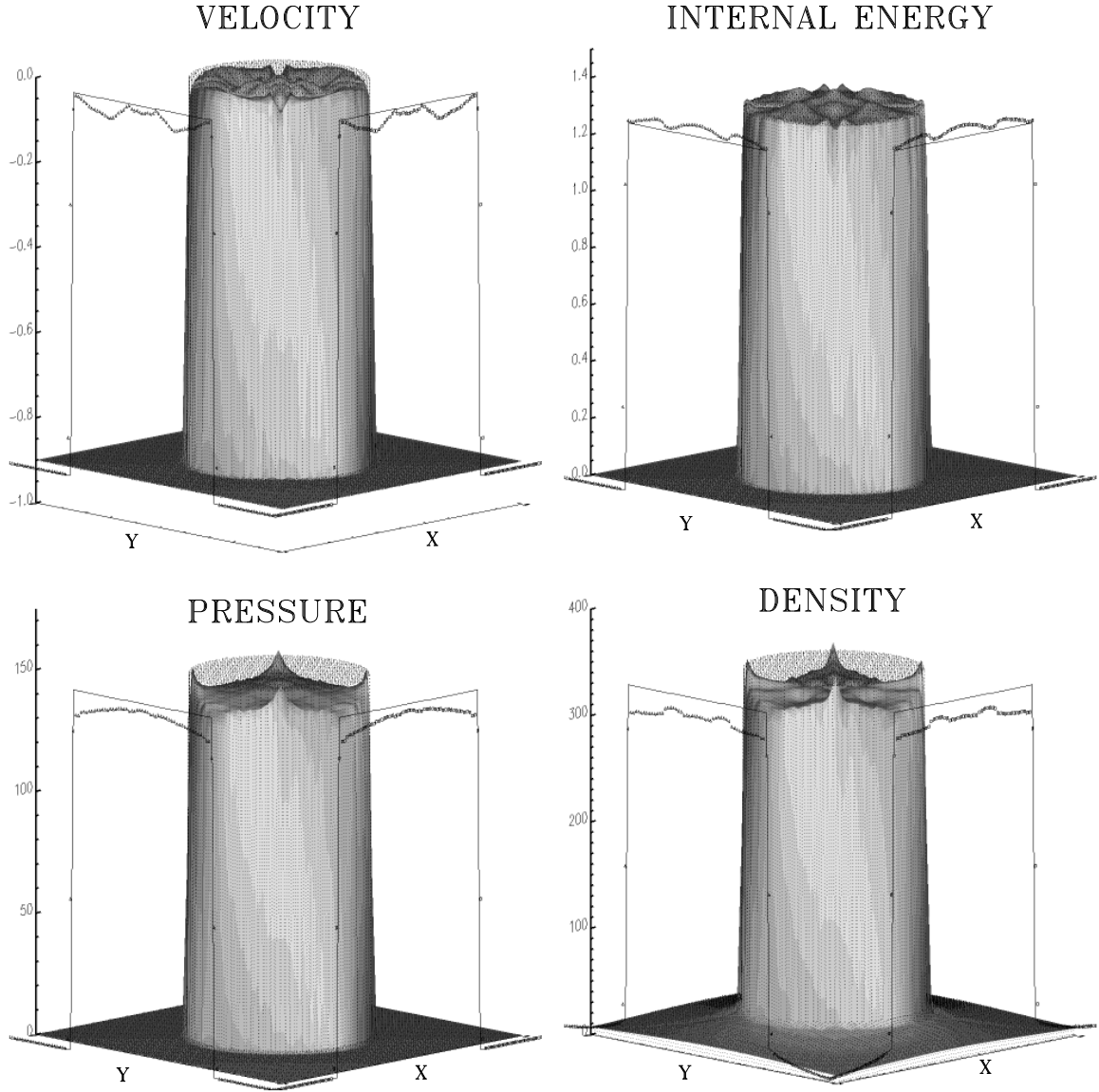}
\caption{Intensity plots of proper rest--mass density, pressure,
specific internal energy and flow velocity over the plane $XY$ at
$z=0$ in the relativistic spherical shock reflection test problem
(RSSR) described in the text, after 2.0 time units. Shaded surfaces
represent the numerical results while dotted surfaces are the exact
solution. One dimensional plots along X and Y axes are projected on
the front sides of the pictures. Symbols inside the one dimensional
plots are numerical values; solid lines represent the exact solution
on the same axis. The test was ran using a CFL equal to 0.2 and a
third--order Runge--Kutta for time integration. \label{fig_ESSR}}
\end{figure}

The initial setup consists of a spherical inflow at speed $v_i$ (which
might be ultrarelativistic) colliding at the centre of symmetry of a
sphere of radius unity. For a hydrodynamic code in Cartesian
coordinates this is a 3D test problem, which allows one to evaluate
the directional splitting technique as well as the symmetry properties
of the algorithm.  Figure \ref{fig_ESSR} shows the numerical results
for $v_i = 0.9c$ on a grid of $101^3$ zones at $t = 2.0$. The shock
capturing properties of GENESIS, which we have already demonstrated in
1D, are retained in this genuine multidimensional case. Two or three
zones are required to handle the shock wave. The pressure and proper
rest--mass density have global relative errors of about $12\%$ and $8\%$
respectively.

\placefigure{fig_ESSR}

Ultrarelativistic flows have been explored by increasing the inflow
Lorentz factor. Table \ref{tab_err2.essr} gives the growth of the
relative global errors (${\epsilon}_{\mbox{\small rel}} =
{\epsilon}_{\mbox{\small abs}} / \left( \sum_{i,j,k}\, \left| {\bf
w}({\bf x}_{i,j,k},t_n) \right|\, \Delta x_j \Delta y_j \Delta z_k
\right)$) on a fixed grid size of $81^3$ zones for $v_i$ in the range
$0.9c$ to $0.999999c$ (the latter inflow velocity corresponding to a
Lorentz factor $W \approx 707$). The relative global errors are
acceptable (considering the inherent difficulty of the test and the
resolution of the experiments) and do not grow dramatically with the
Lorentz factor. The observed growth can be explained by the fact that
the errors are dominated by the shock region and that the shock
strength increases with the Lorentz factor.

  The CFL factors used in the last two tests of this series are
unusually small ($0.019$ and $0.005$) 
which is due to the strength of the shock and the relatively small
grid resolution (compared with the 1D case).  It is noticeable that
for $v_i = 0.999999c$ the errors are considerably larger (last entry
in Table \ref{tab_err2.essr}).  This has two reasons. Firstly, the
global relative errors decrease with time in the RSSR test
problem. Secondly, we could not continue the run with $v_i =
0.999999c$ beyond 1.5 time units, because interaction with the grid
boundaries became severe causing the code to crash.  Hence, $v_i =
0.999999c$ must be considered as the maximum inflow velocity in the
RSSR test problem, which the present code can handle properly (for the
resolution used).  The symmetry properties of the RSSR solution are
very well maintained by GENESIS, even though the number of timesteps
was very large ($> 30000$) in the last two test runs.

  The absolute global errors ($L_1$ norm) and the convergence rates of
the primitive variables at $t=2.0$ are displayed in Table
\ref{tab_err1.essr}. Obviously, the errors are much larger in the 3D
test than in the corresponding 1D one. This can be explained
considering that (i) the grids are coarser than in 1D, and that (ii)
the jumps in pressure and density across the shock are nearly a factor
of 30 larger in the 3D test than in the planar case.
                                                                 
\placetable{tab_err1.essr}

The preservation of the sweep-level symmetry (SLS; see
section \S\ref{symmetry}) is reflected in the symmetry of the one
dimensional profiles in Fig.~\ref{fig_ESSR}. Moreover, a comparison of
the profiles in X and Y direction in Fig.~\ref{fig_ESSR} shows the
capability of the code to maintain the sweep-coherence symmetry (SCS), too.

\section{Code Performance} \label{cperf}

We have parallelized GENESIS in order to run on multiprocessor
computers with shared memory. Appart from the initial setup of
variables, the grid generation and the output, the rest of the program
is organised in a 4-level nested loop. The outermost loop runs from
one to the total number of sub-domains, assigning one sub-domain to
each processor. This procedure allows an almost complete
parallelization of the code employing the corresponding
parallelization directives (see Fig. \ref{flowdiag}).

The MRRP and RSSR tests have been run for different grids on a SGI
Cray-Origin 2000 computer. Tables \ref{tab_perf.gsch} and
\ref{tab_perf.essr} show the total execution time for every run as a
function of the number of CPUs used. We also give the {\it speed up}
factor, defined as the CPU ratio between a one processor run and one
using several processors in parallel.  This factor is a measure of the
degree of parallelization of the code and should ideally be equal to
the number of CPUs used. The tables also contain the execution time
per cell and time iteration (TCI). The TCI for a given number of
processors is nearly independent of the number of computational cells,
and can be used as a time unit to estimate the total execution time
needed in a particular simulation.

According to the data shown in Tables \ref{tab_perf.gsch} and
\ref{tab_perf.essr} the TCI is about $7.6 \,\, 10^{-5}$, $2.1 \,\,
10^{-5}$ and $1.3 \,\, 10^{-5}$ seconds for 1, 4 and 8 processors,
respectively. A significant drop of the performance is noticeable for
a grid of $64^3$ zones due to the phenomenon of {\it cache trashing},
because in this case the dimensions of the 3D matrices are multiples
of the size of cache lines. Hence, different 3D matrices are mapped
into the same set of cache lines, and every time the program needs to
reference a new 3D matrix all cache lines are updated.

\placetable{tab_perf.gsch}

\placetable{tab_perf.essr}

Concerning the speed up factor, it is noticeable from Tables
\ref{tab_perf.gsch} and \ref{tab_perf.essr} that it increases with the
number of grid points, because the 3D nested loops consume a larger
percentage of the total CPU time when the number of grid zones is
larger. The maximum speed up factors are 3.7 and 6.5 for 4 and 8 CPUs,
respectively. We also notice a {\it super linear} behaviour, for the
largest grid, for the MRRP test problem.  As typical 3D simulations are
performed with zone numbers larger than the ones used in the test
runs, we expect to reach even larger speed up factors in these
applications.
 
The number of Mflops (millions of floating point operations per
second) achieved by the code is about 60 on one processor (R10000) of
a SGI Cray--Origin 2000 computer. The theoretical peak speed of such a
processor is 400 Mflops. For comparison, Pen (1998) reports a
performance of 48 Mflops for his 3D adaptive moving mesh classical
hydrodynamic code using a SGI Power Challenge machine with R8000
processors (300 Mflops theoretical peak speed).

Finally, we compare the performance of GENESIS achieved on the PA8000
processor of Hewlett Packard with that obtained on the R10000 processor
of Silicon Graphics. For the comparison we used a HP\,J280 workstation
equipped with a PA8000 processor with a 180 MHz clock and a cache
memory of 512\,Kbytes and a SGI Cray-Origin 2000 equipped with a
R10000 processor with a 195 MHz clock and 4 Mbytes of cache
memory. The test problem selected for the comparison was the
relativistic spherical shock reflection test (RSSR) with an inflow
velocity of $0.9c$. Test runs were done with four different grids.
The resulting execution times per zone and time step (TCI) are
given in Table~\ref{HPvsSGI}.

\placetable{HPvsSGI}

We find that $\mbox{TCI}_{HP} \approx 2 \times \mbox{TCI}_{SGI}$.
From Table~\ref{HPvsSGI} we can infer a general trend. The TCIs
obtained on both machines tend to become similar when the number of
zones increases. Furthermore, the TCI for the HP machine is nearly
independent of the number of zones, while the TCI for the SGI machine
increases with that number. This behaviour may result from the fact
that the problem size always leads to an overflow of the cache memory
on the HP workstation, while this does not generally happen for the
larger cache memory of the SGI machine.

\section{An astrophysical application: axisymmetric jet in 3D}
\label{jet}

\begin{figure}
\plotone{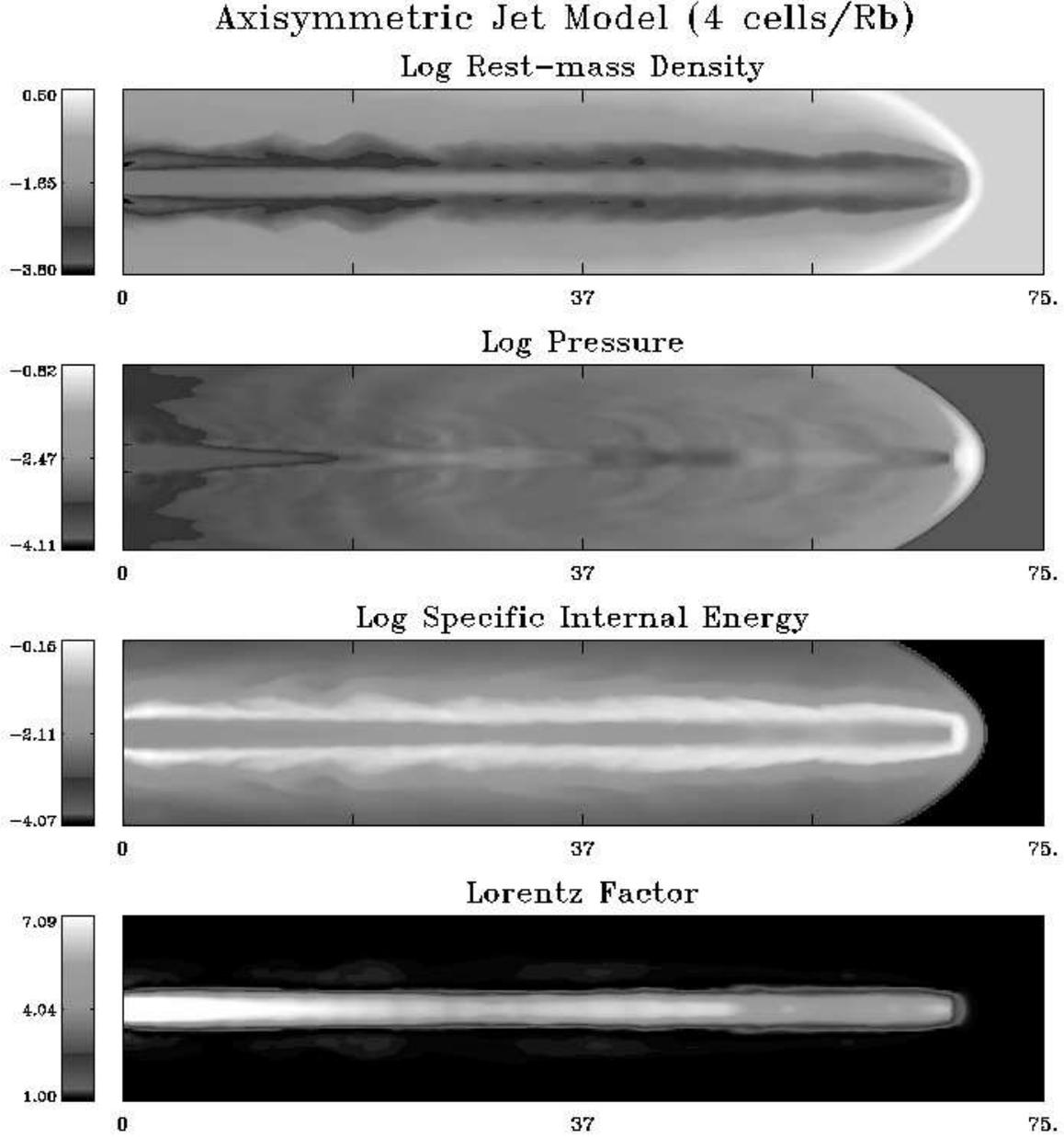}
\caption{Snapshots (from top to bottom) of the proper rest--mass
density distribution, pressure, specific internal energy (all on a
logarithmic scale) and Lorentz factor of the relativistic jet model
discussed in the text ($v_b=0.99c$, ${\cal M}_b=6.0$, ${\eta}=0.01$,
$\gamma=5/3$) after 160 units of time. The resolution is 4
zones$/R_b$.
\label{fig_jet4}}
\end{figure}

\begin{figure}
\plotone{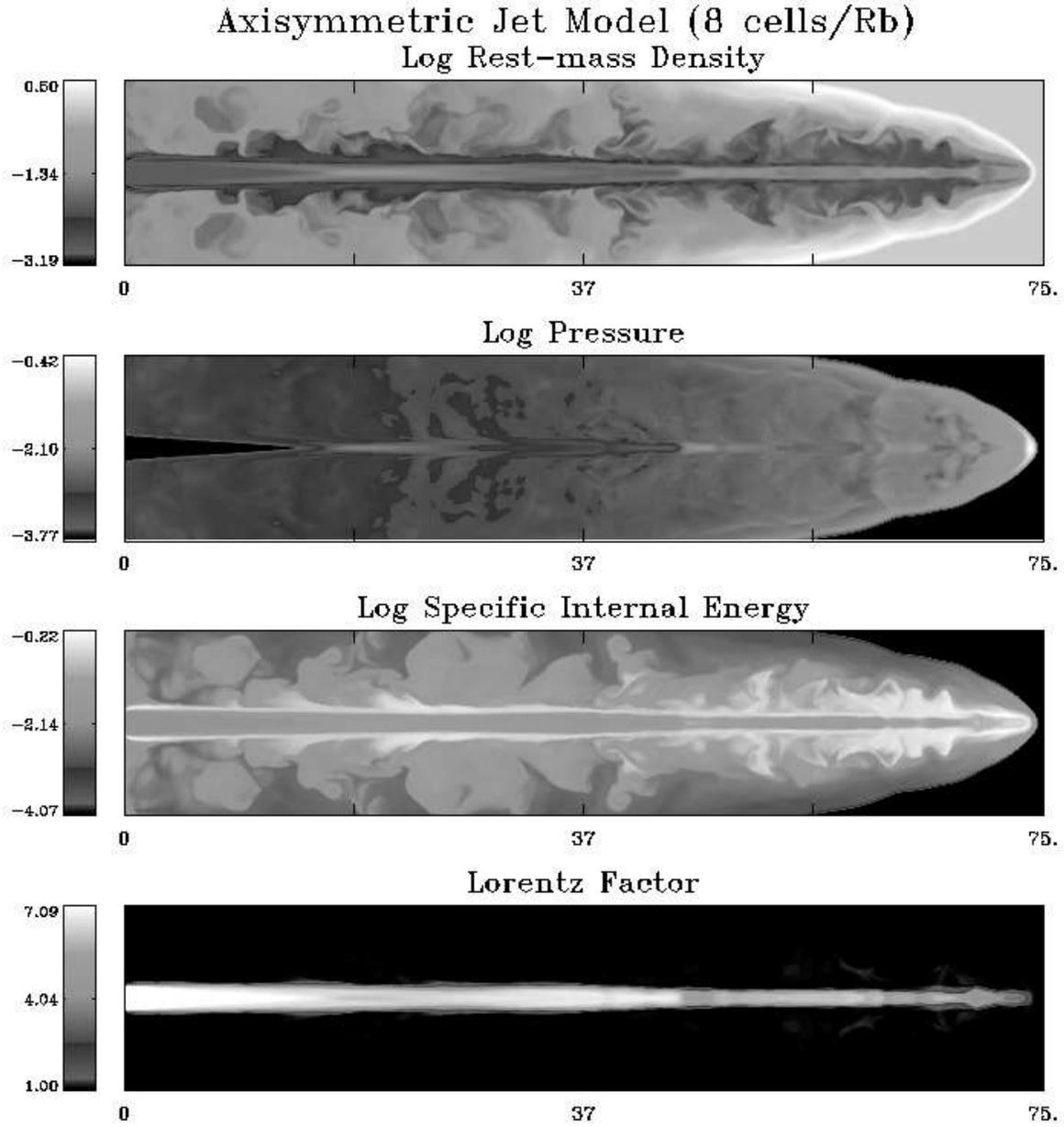}
\caption{Same as Fig. \ref{fig_jet4} but with a resolution of 8 zones$/R_b$.
\label{fig_jet}}
\end{figure}

Next we discuss an astrophysical application computed with GENESIS,
namely the 3D simulation of a relativistic jet propagating through an
homogeneous atmosphere. The properties of the jet are those of model
C2 in Mart\'{\i} et al. (1997). The beam flow velocity, $v_b=0.99c$,
the beam Mach number, ${\cal M}_b=6.0$, and the ratio of the rest-mass
density of the beam and the ambient medium ${\eta}=0.01$.  The ambient
medium is assumed to fill a Cartesian domain (X,Y,Z) with a size of
$15R_b \times 15R_b \times 75R_b$, where $R_b$ is the beam radius.
The jet is injected at $z=0$ in the direction of the positive $z$-axis
through a circular nozzle defined by $x^2+y^2 \le R_b^2$, and is in
pressure equilibrium with the ambient medium.  An ideal gas equation
of state with an adiabatic exponent $\gamma=5/3$ is assumed to
describe both the jet matter and the ambient gas. Two different spatial 
resolutions with 4 and 8 zones per beam radius were used in our calculations
(Figs. \ref{fig_jet4},\ref{fig_jet}).

  In Mart\'{\i} et al. (1997) the simulation was performed in
cylindrical coordinates assuming axial symmetry. The spatial
resolution was 20 zones per beam radius both in the axial and radial
directions. It is well known that the propagation of a supersonic jet
is governed by the interaction of jet matter with the ambient medium,
which produces a bow shock in the ambient medium and an envelope
surrounding the central beam (the {\it cocoon}, in the Blandford \&
Rees 1974 model). The cocoon contains jet material deflected backward
at the head of the jet.  In the case of highly supersonic jets,
discussed in Mart\'{\i} et al. (1997), extensive, overpressured
cocoons are formed with large vortices of jet matter propagating down
the cocoon/ambient medium interface. The vortices are the result of
Kelvin--Helmholtz instabilities at the interface between the jet and
the shocked ambient medium. The interaction of these vortices with the
central beam causes internal shocks inside the beam. These, in turn,
affect the advance speed of the jet making it highly non--stationary.
The propagation speed of the jet can be estimated from the momentum
transfer between the jet and the ambient medium assuming a one
dimensional flow. For model C2 one obtains an advance speed equal to
$0.42c$, whereas the 2D hydrodynamic simulation presented in
Mart\'{\i} et al. (1997) gives a mean jet advance speed of $0.37c$.
 
  The four panels in Figs. \ref{fig_jet4},\ref{fig_jet} display, from top to 
bottom, the logarithm of the proper rest-mass density, pressure and specific
internal energy and flow Lorentz factor in the plane $x=0$ at $t = 160
R_b/c$, when the jet has propagated about 75\,$R_b$. The analysis of
cross sections of the grid perpendicular to the the jet's direction of
propagation (not shown here) reveals acceptable symmetry of the numerical
simulation, i.e both the SLS and the SCS properties are maintained
(see \S \ref{symmetry}). 

  The gross morphological and dynamical properties of highly supersonic
relativistic jets as inferred from our 3D simulations are qualitatively similar 
to those established in earlier 2D simulations. An extensive, overpressured 
cocoon with pressure about 20 times that in the beam at the injection point is 
found surrounding the jet. The pressure and density at the head of the jet in 
the model with 8 zones$/R_b$ are a factor of 2 larger and 1.3 smaller, 
respectively, than in the 2D calculation. For the model with 4 zones$/R_b$ 
these factors are 1 and 1.3, respectively. In contrast with the model with
4 zones$/R_b$, in which the propagation speed coincides with the 1D estimate,
the larger pressure at the head of the jet in the model with 8 zones$/R_b$ 
causes it to propagate through the ambient medium at a larger speed in the 3D
calculation ($0.47c$ instead of $0.42c$ for the 1D estimate and
$0.37c$ for the 2D simulation) producing a narrower profile of the bow
shock near the head. In all the simulations, the supersonic beam displays
rich internal structure with oblique shocks effectively decelerating
the flow in the beam from a Lorentz factor equal to 7 at the injection
point down to a value of about 4 near the head. Whereas gross morphological 
properties are qualitatively similar in all three simulations, finer jet details 
(e.g., number, size, position and development of turbulent vortices in the 
cocoon) do not agree. However, it has been pointed out before that the fine 
structure is highly dependent on the numerical grid resolution (see, e.g., 
K\"ossl \& M\"uller 1988). In the model with 4 zones/$R_b$ (see Fig. 
\ref{fig_jet4}), the material deflected at the head of the jet forms a thick, 
stable overpressured cocoon surrounding the beam up to the nozzle. Due to the 
small resolution only large vortices develope in the cocoon/external medium 
surface which grow slowly. A turbulent cocoon with smaller vortices growing at 
a faster rate (much more similar to the one obtained in the 2D cylindrical model)
are obtained by doubling the resolution (compare, e.g., the proper rest--mass 
density pannels in Figs. \ref{fig_jet4},\ref{fig_jet}).

\placefigure{fig_jet4}

\placefigure{fig_jet}

\section{Conclusions and Future Developments} \label{conc}

We have described the main features of a novel three dimensional,
high--resolution special relativistic hydrodynamic code GENESIS based
on relativistic Riemann solvers. We have discussed several test
problems involving strong shocks in three dimensions which GENESIS has
passed successfully. The performance of GENESIS on single and
multiprocessor machines (HP J280 and SGI Cray--Origin 2000) has been
investigated. Typical simulations (in double precission) with up to
$7\,10^6$ computational cells can be performed with 1Gbyte of RAM
memory with a performance of $\approx 7\,10^{-5}$\,s of CPU time per
zone and time step (on a SCI Cray-Origin 2000 with a R10000
processor). Currently we are working on a version of GENESIS suited
for massively parallel computers with distributed memory (like, e.g.,
Cray T3E).

GENESIS has been designed to handle highly relativistic flows. Hence,
it is well suited for three dimensional simulations of relativistic
jets. First results will be presented in a separate paper (Aloy et
al. 1998). Further applications envisaged are the simulation of
relativistic outflows from merging compact objects (see, e.g., Ruffert
et.al. 1997), from hypernovae (Paczynski 1998), or collapsars
(MacFadyen \& Woosley 1998). In all these models ultra-relativistic
outflow is thought to occur and to play a crucial role in the
generation of gamma-ray bursts.
  
\begin{center}
\subsection*{ACKNOWLEDGEMENTS}
\end{center}
This work has been supported in part by the Spanish DGICYT (grant
PB94-0973 and Acci\'on Integrada hispano-alemana HA1996-0154). MAA
expresses his gratitude to the Conselleria d'Educaci\'o i Ci\`encia de
la Generalitat Valenciana for a fellowship.  The authors gratefully
acknowledge the collaboration of the User's Support Service of the
Centre Europeu de Paral$\cdot$lelisme de Barcelona (CEPBA).  The
calculations were carried out on a HP J280 and on two SGI Origin 2000,
at CEPBA and at the Centre de Inform\'{a}tica de la Universitat de
Val\`encia.

\appendix
\section{Characteristic fields of the RHD equations}

  Analytical expressions for the spectral decomposition of the three $5
\times 5$ (in 3D) Jacobian matrices ${\bf \cal B}^{i}({\bf U})$ associated to
the fluxes ${\bf F}^{i}({\bf U})$ of system (\ref{F}),
\begin{equation}
{\bf \cal B}^{i}({\bf U}) = \frac{\partial{\bf F}^{i}({\bf U})}
{\partial\bf U \rm}
\label{B}
\end{equation}
have been given by Donat et al. (1998).

  In this Appendix, we explicitly show the eigenvalues and the right and
left eigenvectors coresponding to matrix ${\bf \cal B}^{x}$, whereas the cases 
$y$ and $z$ easily follows from symmetry. The eigenvalues are:
\begin{eqnarray}
 \lambda_{\pm} = \frac{1}{1- {\rm v}^{2} c_s^2}
    \left\{v^{x}(1-c_{s}^{2})  {\pm} c_{s}
         \sqrt{(1-{\rm v}^{2}) [1-v^x v^x-({\rm v}^{2}-v^x v^x)c_{s}^{2}]}
    \right\}
\label{lambdapm}
\end{eqnarray}
\begin{equation}
\lambda_0 = v^x \mbox{\,\,\,\,(triple)}
\label{lambda0}
\end{equation}

\noindent
The following expressions define auxiliary quantities:
\begin{eqnarray}
 {\cal K} \equiv
 {\displaystyle{\frac{\tilde{\kappa}}
 {\tilde{\kappa}-c_s^2}}, \:\:\:\: 
 {\tilde \kappa}= \frac{1}{\rho} 
 \left.\frac{\partial p}{\partial \varepsilon}\right|_{\rho}}, 
 \:\:\:\: 
 {\cal A}_{\pm} \equiv
 {\displaystyle{\frac{1 - v^x v^x}{1 - v^x {\lambda}_{\pm}}}}
\end{eqnarray}
\noindent
A complete set of {\it right--eigenvectors} is,
\begin{eqnarray}
 {\bf r}_{0,1} =
 \left( {\displaystyle{\frac{{\cal K}}{h W}}},
 v^x, v^y, v^z, 1 - {\displaystyle{\frac{{\cal K}}{h W}}} \right)
\end{eqnarray}
\begin{eqnarray}
 {\bf r}_{0,2}=
 \left( W v^y , 2 h W^2 v^x v^y,
 h(1+2 W^2 v^y v^y), 2 h W^2 v^y v^z, 2 h W^2 v^y - W v^y\right)
\end{eqnarray}
\begin{eqnarray}
 {\bf r}_{0,3}=
 \left( W v^z , 2 h W^2 v^x v^z,
 2 h W^2 v^y v^z, h(1+2 W^2 v^z v^z),  2 h W^2 v^z - W v^z\right)
\end{eqnarray}
\begin{eqnarray}
 {\bf r}_{\pm} = (1, h W {\cal A}_{\pm} {\lambda}_{\pm}, h W v^y,
 h W v^z, h W {\cal A}_{\pm} - 1)
\end{eqnarray}

\noindent
The corresponding complete set of {\it left--eigenvectors} is 
\begin{eqnarray*}
 {\bf l}_{0,1} = {\displaystyle{\frac{W}{{\cal K} - 1}} }
 (h - W, W v^x, W v^y, W v^z, -W)
\end{eqnarray*}
\begin{eqnarray*}
 {\bf l}_{0,2} = {\displaystyle{\frac{1}{h (1 - v^x v^x)}} }
 (- v^y, v^x v^y, 1 - v^x v^x, 0, -v^y)
\end{eqnarray*}
\begin{eqnarray*}
 {\bf l}_{0,3} = {\displaystyle{\frac{1}{h (1 - v^x v^x)}} }
 (- v^z, v^x v^z, 0, 1 - v^x v^x, -v^z)
\end{eqnarray*}

\[
{\bf l}_{\mp} = ({\pm} 1){\displaystyle{\frac{h^2}{\Delta}}}
 \left[ \begin{array}{c}
 h W {\cal A}_{\pm} (v^x - {\lambda}_{\pm}) -
 v^x - W^2 ({\rm v}^{2} - v^x v^x) (2 {\cal K} - 1)
 (v^x - {\cal A}_{\pm} {\lambda}_{\pm}) +
 {\cal K} {\cal A}_{\pm} {\lambda}_{\pm}
\\  \\
 1 + W^2 ({\rm v}^{2} - v^x v^x) (2 {\cal K} - 1) (1 - {\cal A}_{\pm}) -
	{\cal K} {\cal A}_{\pm}  \\ \\
	W^2 v^y (2 {\cal K} - 1) {\cal A}_{\pm} (v^x - {\lambda}_{\pm}) \\ \\
	W^2 v^z (2 {\cal K} - 1) {\cal A}_{\pm} (v^x - {\lambda}_{\pm}) \\ \\
	- v^x - W^2 ({\rm v}^{2} - v^x v^x) (2 {\cal K} - 1)
	(v^x - {\cal A}_{\pm} {\lambda}_{\pm}) +
	{\cal K} {\cal A}_{\pm} {\lambda}_{\pm}
	\end{array} \right]
\]
where $\Delta$ is the determinant of the matrix of right-eigenvectors.
\begin{eqnarray}
 \Delta = h^3 W ({\cal K} - 1) (1 - v^x v^x)
 ({\cal A}_{+} {\lambda}_{+} - {\cal A}_{-} {\lambda}_{-})
\end{eqnarray}
For an ideal gas equation of state it can be proven that ${\cal K}$ is
always greater than one (in fact ${\cal K}$ = $h$), and $\Delta$ is
different from zero ($\mid v^x \mid < 1$).
\newpage

\section{Numerical fluxes and disipation terms in Marquina's flux 
formula} \label{visco}

  Given a cell interface, the numerical fluxes accross such interface
are given, in our modified version of Marquina's flux formula, by
\begin{eqnarray}
{\bf \tilde{F}} =  \frac{1}{2} 
\left 
({\bf F}^{L}+{\bf F}^{R} +
 {\bf Q} \right).
\end{eqnarray}
In the above expression, ${\bf F}^{L}$ and ${\bf F}^{R}$ are the
fluxes at the left and right side of the interface and ${\bf Q}$ is a
five--vector containing the numerical viscosity terms which is
calculated according to ${\bf Q} = {\bf Q}^R - {\bf Q}^L$.

Quantities ${\bf Q}^S$ ($S=L,R$) are written as a sum involving the
projectors onto each eigenspace (i.e., direct products of left and
right eigenvectors) and the eigenvalues of the corresponding Jacobian
matrix
\begin{eqnarray}
Q^S_j=\sum_{i,k=1}^5 |\lambda_i|_{\mbox{max(L,R)}} \,\, r^S_{ji} \,\, l^S_{ik} 
\,\, U^S_k, \,\,\,\,\,\,\, (j=1,\ldots,5).
\end{eqnarray}
\noindent
Quantities $|\lambda_i|_{\mbox{max(L,R)}}$ ($i=1,\ldots,5$) are the
maximum of the modulus of the two corresponding eigenvalues at the
left and right side of the interface, whereas $r^S_{ij}$ ($l^S_{ij}$)
with $i,j=1,\ldots,5$, refer to the $i$ ($j$) component of the right
(left) eigenvector $j$ ($i$). In this expression, subscripts ${1,
\ldots,5}$ correspond to ${-,0,0,0,+}$ as defined in Appendix~A.
$U^S_k$ are the components of the vector of unknowns. The superscript
$S$ indicates that the various qunatities are calculated at each side
of the interface in terms of the reconstructed variables.

  Omitting the superscript $S$ (the expressions are identical for the
left and right side of the interface) one derives the following
analytical formulas for $Q_j^S$:
\begin{eqnarray*}
Q_1 &=& \frac{h}{\Delta} \left\{ M \left[ {\cal A}_{-} {\Omega}_{+} - 
                                          {\cal A}_{+} {\Omega}_{-} \right] + 
p(\tilde{\lambda}_{+}\tilde{l}_{+}-\tilde{\lambda}_{-}\tilde{l}_{-}) \right\} + \\
& & \tilde{\lambda}_{0}p\frac{W}{h}\left\{ \frac{{\cal K}}{{\cal K} - 1} + \frac{v_t^2+v_{tt}^2}
{1-v_n^2} \right \} \\
Q_2 &=& \frac{h^2W}{\Delta} \left\{ M{\cal A}_{+}{\cal A}_{-} \left[ 
{\Omega}_{+}\lambda_{+}-{\Omega}_{-}\lambda_{-}
\right] + 
p(\tilde{\lambda}_{+}\lambda_{+}{\cal A}_{+}\tilde{l}_{+}-
         \tilde{\lambda}_{-}\lambda_{-}{\cal A}_{-}\tilde{l}_{-}) \right\} + \\
& & \tilde{\lambda}_{0}pW^2v_n \left\{ \frac{1}{{\cal K} - 1} + 2\frac{v_t^2+v_{tt}^2}{1-v_n^2} 
\right \} \\
Q_3 &=& \frac{h^2W}{\Delta}v_t  \left\{ M \left[ 
{\Omega}_{+}{\cal A}_{-} -
{\Omega}_{-}{\cal A}_{+} \right] + 
p(\tilde{\lambda}_{+}\tilde{l}_{+}-\tilde{\lambda}_{-}\tilde{l}_{-}) \right\} + \\
& & \tilde{\lambda}_{0}p \left\{ \frac{W^2}{{\cal K} - 1} + \frac{1+2W^2(v_t^2+v_{tt}^2)}{1-v_n^2} 
\right \} \\
Q_4 &=& \frac{h^2W}{\Delta}v_{tt} \left\{ M \left[ 
{\Omega}_{+}{\cal A}_{-}-
{\Omega}_{-}{\cal A}_{+} \right] + 
p(\tilde{\lambda}_{+}\tilde{l}_{+}-\tilde{\lambda}_{-}\tilde{l}_{-}) \right\} + \\
& & \tilde{\lambda}_{0}p \left\{ \frac{W^2}{{\cal K} - 1} + \frac{1+2W^2(v_t^2+v_{tt}^2)}{1-v_n^2} 
\right \} \\
Q_5 &=& \frac{h}{\Delta} \left\{ M \left[ {\cal A}_{-} {\Omega}_{+}{\cal D}_{+}-
                     {\cal A}_{+} {\Omega}_{-}{\cal D}_{-} \right] + 
p[\tilde{\lambda}_{+}\tilde{l}_{+}{\cal D}_{+} -
         \tilde{\lambda}_{-}\tilde{l}_{-}{\cal D}_{-}] \right\} + \\
& & \tilde{\lambda}_{0}p\frac{W}{h}\left\{ \frac{hW-{\cal K}}{{\cal K} - 1} + 
\frac{(2hW-1)(v_t^2+v_{tt}^2)}{1-v_n^2} \right \},
\label{viscos}
\end{eqnarray*}
with 
$M=\rho hW^2({\cal K}-1)$, 
${\Omega}_{\pm}= \tilde{\lambda}_{\pm}(v_n - \lambda_{\mp})$, 
${\cal D}_{\pm}= hW{\cal A}_{\pm}-1$, 
$\tilde{\lambda}_{+}=|\lambda_5|_{\mbox{max(L,R)}}$,
$\tilde{\lambda}_{0}=|\lambda_2|_{\mbox{max(L,R)}}$ and
$\tilde{\lambda}_{-}=|\lambda_1|_{\mbox{max(L,R)}}$. 
Quantities $v_n$, $v_t$ y $v_{tt}$ denote the velocities normal and
parallel to the interface at which the numerical flux is to be
computed, and $\tilde{l}_{\pm}$ are quantities proportional to the
fifth component of the left eigenvectors ${\bf l}_{+}$ and ${\bf
l}_{-}$ given by:
\begin{eqnarray*}
{\tilde l}_{\pm} = 
{\pm}\left \{ -v^x -W^2({\rm v}^{2}-v^x v^x)(2{\cal{K}}-1)(v^x - {\cal{A}_{\pm}}\lambda_{\pm})
+{\cal{K} \cal{A}_{\pm}} \lambda_{\pm} \right\}
\end{eqnarray*}

\section{Explicit algorithm to recover primitive variables}

  In any RHD code evolving the conserved quantities Eq.~(\ref{U}) in
time, the variables $\{p,v^1,v^2,v^3,\rho,\varepsilon\}$ have to be
computed from the conserved quantities at least once per time step. In
GENESIS this is achieved using Eqs.~(\ref{D})--(\ref{tau}) and the
equation of state.  For an ideal gas equation of state with constant
$\gamma$, this implies to find the root of the function
\begin{equation}
f(p)=(\gamma - 1) \rho_\ast \varepsilon_\ast - p
\label{16}
\end{equation}
with $\rho_\ast$ and $\varepsilon_\ast$ given by
\begin{equation}
\rho_\ast=\frac{D}{W_\ast}
\label{17}
\end{equation}
and
\begin{equation}
\varepsilon_\ast = \frac{\tau + D\,(1 - W_\ast) +
p\,(1 - W_\ast^2)}{D\,W_\ast} \, ,
\label{18}
\end{equation}
where
\begin{equation}
W_\ast = \frac{1}{{\displaystyle \sqrt{1- {\bf v}_\ast \cdot {\bf v}_\ast}}}  
\, ,
\label{19}
\end{equation}
and 
\begin{equation}
{\bf v}_\ast = \frac{\bf S}{\tau + D + p} \, .
\label{rmv}
\end{equation}

  The zero of $f(p)$ in the physically allowed domain $p \in \,]p_{\rm
min}, \infty [$ determines the pressure. The monotonicity of $f(p)$ in
that domain ensures the uniqueness of the solution. The lower bound of
the physically allowed domain, $p_{\rm min}$, defined by
\begin{equation}
p_{\rm min} = |{\bf S}| - \tau - D,
\label{20}
\end{equation}
is obtained from Eq.~(\ref{rmv}) taking into account that (in our
units) $|{\bf v}| \leq 1$.  Knowing $p$, Eq.~(\ref{rmv}) then directly
gives ${\bf v}$, while the remaining state quantities are
straightforwardly computed from Eqs.~(\ref{D})--(\ref{tau}) and the
definition of the Lorentz factor.

  In GENESIS, the solution of $f(p) = 0$ is obtained by means of a 
Newton--Raphson iteration in which the derivative of $f$, $f^\prime$, is
approximated by 
\begin{equation}
f^\prime = |{\bf v}_\ast|^2 c_{s \ast}^2 - 1,
\end{equation}
where $c_{s \ast}$ is the sound speed given by
\begin{equation}
c_{s \ast} = \sqrt{\frac{(\gamma -1)\gamma \varepsilon_\ast}{1+\gamma
\varepsilon_\ast}}.
\end{equation} 
This approximation tends to the exact derivative when the solution is
approached. On the other hand, it easily allows one to extend the
present algorithm to general equations of state.

\newpage

\begin{deluxetable}{cccc}

\tablecaption{Absolute global errors ($L_1$ norm) of the primitive
variables for the mildly relativistic Riemann test problem (MRRP) for
different grids at $t=0.5$. As the errors are dominated by those zones
located inside the shock and as the grid resolution is still poor even
on the finest grid, we have repeated every calculation four times
varying $t$ within an interval $t \pm \delta t$ ($\delta t$ being of
the order of one Courant time) and calculated the mean errors. In
parentheses we give the standard root mean square deviation of the
errors (${\sigma}_{\mbox{n-1}}$). \label{tab_err1.schn}}

\tablewidth{0pt}

\tablehead{
\colhead{Cells}  & \colhead{Pressure} & \colhead{Density} & 
\colhead{Velocity}
}
\startdata 
$40^3$ & 8.0(2.0)E--2 & 1.1(0.3)E--1 & 0.9(0.4)E--2 \nl
$60^3$ & 5.2(0.4)E--2 & 9.8(0.8)E--2 & 1.1(0.3)E--2 \nl
$80^3$ & 4.5(0.2)E--2 & 9.2(0.5)E--2 & 1.1(0.1)E--2 \nl
$100^3$& 3.7(0.4)E--2 & 7.0(0.9)E--2 & 7.0(2.0)E--3 \nl
$150^3$& 2.5(0.2)E--2 & 4.8(0.7)E--2 & 5.0(2.0)E--3 \nl
\enddata
\end{deluxetable}

\begin{deluxetable}{ccccccc}
\tablecaption{Absolute global errors ($L_1$ norm) of the primitive
variables (columns 2-4) and the corresponding convergence rates
(columns 5-7) for the relativistic planar shock reflection test
problem (RPSR) for different grids at $t=2.0$.  The test runs have
been performed with a Courant number equal to 0.1 and the third order
accurate Runge-Kutta time integration method (RK3). In parenthesis we
give the standard root mean square deviation of the errors (see also
Table \ref{tab_err1.schn}).
\label{tab_err1.rssr}}
\tablewidth{0pt}
\tablehead{
\colhead{Cells} & \colhead{Pressure}  & \colhead{Density} & 
\colhead{Velocity} & \colhead{$r_{\rho}$} & \colhead{$r_{p}$} & 
\colhead{$r_{v}$} 
}
\startdata
$101$ & 19.3(0.3)E+0  & 290.8(0.4)E--2 &  2.4(0.1)E--2 &            &         &         \nl
$201$ & 10.8(0.2)E+0  & 147.2(0.7)E--2 & 10.1(0.4)E--3 & 0.99       &  0.84   &  1.26   \nl
$401$ & 49.2(0.7)E--1 &  85.0(1.0)E--2 & 92.8(0.8)E--4 & 0.80       &  1.14   &  0.14   \nl 
$801$ & 25.2(0.2)E--1 &  37.3(0.1)E--2 &  3.4(0.1)E--3 & 1.19       &  0.97   &  1.44   \nl
$1601$& 13.8(0.1)E--1 & 187.4(0.7)E--3 & 17.3(0.4)E--4 & 0.99       &  0.87   &  0.98   \nl
\enddata
\end{deluxetable}

\begin{deluxetable}{cccc}
\tablecaption{Relative global errors ($L_1$ norm) of the primitive
variables for the planar shock reflection test problem (RPSR) on a
grid of 401 zones at $t=2.0$. The quantity $\nu$ is defined as $\nu =
1 - v_i$.  The test runs have been performed with a Courant number
equal to 0.1 and the third order accurate Runge-Kutta time integration
method (RK3). In parenthesis we give the standard root mean square
deviation of the errors (see also Table \ref{tab_err1.schn}).
\label{tab_err4.rssr}}
\tablewidth{0pt}
\tablehead{
\colhead{$\nu$} & \colhead{Pressure} & \colhead{Density} & \colhead{Velocity}
}
\startdata
 $10^{-1}$  &  90.7(0.5)E--4 &  96.6(0.5)E--4 & 80.3(0.5)E--4 \nl
 $10^{-3}$  &  58.0(0.8)E--4 &  72.0(0.8)E--4 & 12.6(0.1)E--3 \nl 
 $10^{-5}$  & 100.3(0.5)E--5 &  79.3(0.5)E--4 & 72.0(0.8)E--4 \nl 
 $10^{-7}$  &  61.0(0.8)E--4 &  93.0(0.1)E--4 & 85.6(0.1)E--4 \nl
 $10^{-9}$  &  65.2(0.1)E--4 & 103.0(0.1)E--4 & 81.3(0.5)E--4 \nl
 $10^{-11}$ & 141.0(0.1)E--5 & 340.1(0.1)E--4 &325.7(0.5)E--5 \nl
 \enddata
\end{deluxetable}

\begin{deluxetable}{ccccccc}
\tablecaption{Absolute global errors ($L_1$ norm) and convergence
rates of the primitive variables for the relativistic spherical shock
reflection test problem (RSSR) for different grids at $t=2.0$. The
test runs have been performed with a Courant number equal to 0.1 and
the third order accurate Runge-Kutta time integration method (RK3). In
parenthesis we give the standard root mean square deviation of the
errors (see also Table \ref{tab_err1.schn}).
\label{tab_err1.essr}}
\tablewidth{0pt}
\tablehead{
\colhead{Cells} & \colhead{Pressure} & \colhead{Density} & \colhead{Velocity} & 
\colhead{$r_{\rho}$} & \colhead{$r_{p}$} & \colhead{$r_{v}$} 
}
\startdata
$41^3$ & 11.8(0.2)E+0  & 30.3(0.4)E+0 & 80.0(3.0)E--3  &      &      &      \nl
$61^3$ & 76.5(0.7)E--1 & 20.1(0.2)E+0 & 55.8(0.6)E--3  & 1.09 & 1.03 & 0.91 \nl 
$81^3$ & 57.5(0.8)E--1 & 15.5(0.2)E+0 & 41.0(0.8)E--3  & 1.01 & 0.92 & 1.09 \nl
$101^3$& 45.2(0.8)E--1 & 12.5(0.1)E+0 & 32.4(0.5)E--3  & 0.99 & 0.97 & 1.07 \nl
\enddata
\end{deluxetable}

\begin{deluxetable}{cccc}
\tablecaption{Growth of relative global errors of the primitive
variables for the relativistic spherical shock reflection test problem
(RSSR) for different inflow velocities at $t=2.0$. The four test runs
have been performed with RK3 and Courant numbers 0.1, 0.1, 0.019 and
0.005, respectively. The quantity $\nu$ has the same meaning as in
Table \ref{tab_err4.rssr}.
\label{tab_err2.essr}}
\tablewidth{0pt}
\tablehead{
\colhead{$\nu$} & \colhead{Pressure $(\%)$} & \colhead{Density $(\%)$} & 
\colhead{Velocity $(\%)$}}
\startdata
 $10^{-1}$  &   15.8 &  10.5 &  0.82  \nl
 $10^{-3}$  &   19.9 &  22.1 &  3.07  \nl
 $10^{-5}$  &   22.1 &  27.8 &  3.89  \nl 
 $10^{-6} \tablenotemark{(a)}$  &   32.2 &  39.1 &  1.91  \nl  
\enddata
\tablenotetext{(a)}{The run time for this test is $1.5$.}
\end{deluxetable}

\begin{deluxetable}{cccccccc}
\tablecaption{Performance of GENESIS for the mildly
relativistic Riemann test problem (MRRP) on different grids. The test
runs are stopped at $t=0.5$, and are performed with a Courant number
equal to 0.8 and the second order accurate Runge-Kutta time
integration (RK2) method.  Times are measured in seconds on a SGI
Cray--Origin 2000.  The last column displays the number of Mflops per
processor and the total number of Mflops. One notices that the
efficiency per processor in parallel mode (Speed Up/CPUs) multiplied
by the number of Mflops in sequential mode is equal to the number of
Mflops in parallel mode. Megaflops are calculated using SGI's {\it
Perfex Tool}.
\label{tab_perf.gsch}}
\tablewidth{0pt}
\tablehead{
\colhead{\# cells} & \colhead{\# CPUs} & \colhead{Time} & \colhead{Speed up} & 
\colhead{\# iter} & \colhead{TCI} & \colhead{Mflops}
}
\startdata
 $44^3$ &   1    & 3.91E2 &           &    86  & 5.34E--5 &  64.73/------ \nl
        &   4    & 1.13E2 &   3.48    &        & 1.54E--5 &  59.03/236.11 \nl
        &   8    & 6.02E1 &   6.50    &        & 8.22E--6 &  58.58/468.65 \nl
\hline
 $64^3$ &   1    & 3.85E3 &          &   118  & 1.24E--4 &  30.05/------ \nl
        &   4    & 1.84E3 &   2.09   &        & 5.94E--5 &  16.25/65.00 \nl
        &   8    & 1.49E3 &   2.59   &        & 4.81E--5 &  10.46/83.70 \nl
\hline
 $84^3$ &   1    & 5.56E3 &          &   150  & 6.26E--5 &  62.04/------ \nl
        &   4    & 1.56E3 &   3.57   &        & 1.75E--5 &  56.77/227.08  \nl
        &   8    & 8.46E2 &   6.58   &        & 9.52E--6 &  53.99/431.88  \nl
\hline
 $104^3$&   1    & 1.31E4 &          &   183  & 6.35E--5 &  62.72/------ \nl
        &   4    & 3.66E3 &   3.57   &        & 1.78E--5 &  57.02/228.08 \nl
        &   8    & 2.36E3 &   5.54   &        & 1.15E--5 &  45.47/363.75 \nl 
\hline
 $154^3$&   1    & 8.94E4 &          &   265  & 9.23E--5 &  45.12/------ \nl
        &   4    & 1.84E4 &   4.87   &        & 1.90E--5 &  54.92/219.68 \nl
        &   8    & 1.15E4 &   7.80   &        & 1.18E--5 &  44.74/357.91 \nl 
        &   16   & 7.39E3 &  12.09   &        & 7.64E--6 &  35.90/574.41 \nl 
\enddata
\end{deluxetable}

\begin{deluxetable}{ccccccc}
\tablecaption{Performance of GENESIS for the relativistic
spherical shock reflection test problem (RSSR) on different grids. The
test runs are stopped at $t=2.0$, and are performed with a Courant
number varying from 0.8 ($45^3$ grid) to 0.2 ($105^3$ grid). The third
order accurate Runge-Kutta time integration (RK3) method has been
used.  Times are measured in seconds on a SGI Cray--Origin 2000.  The
last column displays the number of Mflops per processor and the total
number of Mflops. One notices that the efficiency per processor in
parallel mode (Speed Up/CPUs) multiplied by the number of Mflops in
sequential mode is equal to the number of Mflops in parallel
mode. Megaflops are calculated using SGI's {\it Perfex Tool}.
\label{tab_perf.essr}}

\tablewidth{0pt}
\tablehead{
\colhead{\# cells} & \colhead{\# CPUs} & \colhead{Time} & \colhead{Speed up} & 
\colhead{\# iter} & \colhead{TCI} & \colhead{Mflops}
}
\startdata
 $45^3$ &   1    & 8.73E2 &          &  114   & 8.40E--5 &  62.53/------ \nl
        &   4    & 2.53E2 &   3.45   &        & 2.44E--5 &  56.88/225.67 \nl
        &   8    & 1.51E2 &   5.78   &        & 1.45E--5 &  50.15/401.21 \nl
\hline
 $65^3$ &   1    & 4.94E3 &          &  198   & 9.08E--5 &  62.27/------ \nl
        &   4    & 1.50E3 &   3.29   &        & 2.76E--5 &  52.88/211.54 \nl
        &   8    & 1.10E3 &   4.49   &        & 2.02E--5 &  37.54/300.28 \nl
\hline
 $85^3$ &   1    & 2.15E4 &          &  369   & 9.49E--5 &  61.76/------ \nl
        &   4    & 6.13E3 &   3.51   &        & 2.71E--5 &  55.40/221.60 \nl
        &   8    & 3.41E3 &   6.30   &        & 1.51E--5 &  51.35/410.79 \nl
\hline
 $105^3$&   1    & 9.92E4 &          &  890   & 9.63E--5 &  62.09/------ \nl
        &   4    & 2.78E4 &   3.57   &        & 2.70E--5 &  56.41/225.65 \nl
        &   8    & 1.57E3 &   6.31   &        & 1.53E--5 &  48.97/391.79 \nl 
\hline
\enddata
\end{deluxetable}

\begin{deluxetable}{ccccccc}
\tablecaption{Performance of GENESIS for the relativistic
spherical shock reflection test (RSSR) on different grids and
machines.
\label{HPvsSGI}}
\tablewidth{0pt}
\tablehead{
\colhead{\# cells} & \colhead{Machine} & \colhead{Time} & 
\colhead{\# iter} & \colhead{TCI}
}
\startdata
  $45^3$ &   SGI   & 8.73E2 &   114   & 8.40E-5 \nl
         &   HP    & 2.11E3 &   101   & 2.29E-4 \nl
  $65^3$ &   SGI   & 4.94E3 &   198   & 9.08E-5 \nl
         &   HP    & 1.33E4 &   220   & 2.20E-4 \nl
  $85^3$ &   SGI   & 2.15E4 &   369   & 9.49E-5 \nl
         &   HP    & 4.76E4 &   357   & 2.17E-4 \nl
 $105^3$ &   SGI   & 9.92E4 &   890   & 9.63E-5 \nl
         &   HP    & 2.16E5 &   879   & 2.12E-4 \nl  
\enddata
\end{deluxetable}

\newpage

\end{document}